\documentclass{ws-procs9x6}            

\usepackage{epsfig}
\usepackage{ graphicx } 
\usepackage{ bm } 
\usepackage{ color }

\newcommand{\ua}{\uparrow}
\newcommand{\da}{\downarrow}

\newcommand{\be}{\begin{equation}}
\newcommand{\ee}{\end{equation}}
\newcommand{\bea}{\begin{eqnarray}}
\newcommand{\eea}{\end{eqnarray}}

\newcommand{\veps}{\varepsilon}
\newcommand{\vk}{{\boldsymbol k}}

\begin{document}
\title{Dynamics of quenched topological edge modes}

\author{P. D. Sacramento$^*$}

\address{Departamento de F\'{\i}sica and CeFEMA,\\ Instituto Superior T\'ecnico,
Universidade de Lisboa,\\
Av. Rovisco Pais, 1049-001 Lisboa, Portugal\\
$^*$E-mail: pdss@cefema-gt.tecnico.ulisboa.pt}

\begin{abstract}
A characteristic feature of topological systems is the presence
of robust gapless edge states. In this work the effect of time-dependent
perturbations on the edge states is considered. Specifically we consider
perturbations that can be understood as changes of the parameters of
the Hamiltonian. These changes may be sudden or carried out at a fixed rate.
In general, the edge modes decay in the thermodynamic limit, but for finite
systems a revival time is found that scales with the system size.
The dynamics of fermionic edge modes and Majorana modes are compared.
The effect of periodic perturbations is also referred allowing the
appearance of edge modes out of a topologically trivial phase.

\end{abstract}


\bodymatter

\section{Sudden quantum quenches}

An example of a time-dependent transformation of the Hamiltonian
is a sudden change of its parameters.
Let us consider an Hamiltonian defined by an initial set of
parameters $\xi_0$ for times $t<t_0$.
The single-particle eigenstates of the Hamiltonian are given by
\be
H(\xi_0) |\psi_{m_0}(\xi_0)\rangle = E_{m_0}(\xi_0)|\psi_{m_0}(\xi_0)\rangle ,
\ee
where $m_0$ are the quantum numbers.
At time $t=t_0$ a sudden transformation of the parameters is performed,
$\xi_0 \rightarrow \xi_1$. The Hamiltonian eigenstates transform to
\be
H(\xi_1) |\psi_{m_1}(\xi_1)\rangle = E_{m_1}(\xi_1)|\psi_{m_1}(\xi_1)\rangle .
\ee
After this sudden quench
the system will evolve in time under the influence of a different Hamiltonian.
The time evolution of a single-particle state, with quantum number $m_0$, is given by
\bea
|\psi_{m_0}^I(t) \rangle &=& \sum_{m_1} e^{-i E_{m_1}(\xi_1) (t-t_0)} 
\nonumber \\
& & |\psi_{m_1}(\xi_1) \rangle
\langle \psi_{m_1}(\xi_1)|\psi_{m_0}(\xi_0) \rangle
\eea
for times $t \geq t_0$. 
The survival probability of some initial state $|\psi_{m_0}(\xi_0) \rangle$ is
defined by
\be
P_{m_0}(t)=|\langle \psi_{m_0}(\xi_0) |\psi_{m_0}^I(t) \rangle |^2 .
\ee

We will be interested in the fate of single particle states after a quantum
quench across the phase diagram. We consider a subspace of one excitation such that
the total Hamiltonian is given by the ground state energy plus one excited state
and assume we remain in the one excitation subspace after the quench.
In this work only unitary evolution of single-particle states is considered
and effects of dissipation are neglected.

We may as well consider further quenches defined in a sequence of times and sets
of parameters as $t_0<t_1<t_2<t_3<\cdots$ and $\xi_0, \xi_1,\xi_2,\xi_3,\cdots$,
respectively. These intervals define regions as
$I (t_0 \leq t <t_1), II (t_1 \leq t < t_2), III (t_2 \leq t <t_3), \cdots$.
The case of a single quench is clearly obtained taking $t_1 \rightarrow \infty$, and
so on for further quenches
($t_0=0$ is chosen hereafter).

Consider now a case for which we have two quenches in succession. In this case we have that
the evolution of the initial state with quantum number $m_0$ is
\bea
|\psi_{m_0}^{II}(t) \rangle &=& e^{-i H(\xi_2)(t-t_1)} |\psi_{m_0}^I(t_1) \rangle \nonumber \\
&=& \sum_{m_2} e^{-i E_{m_2}(\xi_2)(t-t_1)} 
\nonumber \\
& & |\psi_{m_2}(\xi_2) \rangle \langle \psi_{m_2}(\xi_2)|
\psi_{m_0}^I(t_1) \rangle \nonumber \\
&=& \sum_{m_2} \sum_{m_1} e^{-i E_{m_2}(\xi_2)(t-t_1)} e^{-i E_{m_1}(\xi_1)t_1} 
\nonumber \\
& & |\psi_{m_2}(\xi_2) \rangle
 \langle \psi_{m_2}(\xi_2)|\psi_{m_1}(\xi_1) \rangle \langle \psi_{m_1}(\xi_1)|\psi_{m_0}(\xi_0) \rangle
\nonumber \\
& & 
\eea

Choosing $\xi_2=\xi_0$ we get that for $t_1 \leq t<\infty$ ($t_2 \rightarrow \infty$)
the overlap with an initial state, $n_0$, is given by
\bea
\langle \psi_{n_0}(\xi_0)|\psi_{m_0}^{II}(t) \rangle &=& \sum_{m_1} e^{-i E_{n_0}(\xi_0)(t-t_1)}
e^{-i E_{m_1}(\xi_1) t_1} \nonumber \\ 
& & \langle \psi_{n_0}(\xi_0)|\psi_{m_1}(\xi_1) \rangle \langle \psi_{m_1}(\xi_1)|\psi_{m_0}(\xi_0) \rangle
\nonumber \\
& & 
\eea
Therefore, the probability to find a projection to an initial state, $n_0$, given that the
initial state is $m_0$ is given by
\bea
P_{n_0m_0}(t) &=& |\langle \psi_{n_0}(\xi_0)|\psi_{m_0}^{II}(t) \rangle |^2 \nonumber \\
&=& | \sum_{m_1} e^{-i E_{m_1}(\xi_1) t_1} \nonumber \\
& & \langle \psi_{n_0}(\xi_0)|\psi_{m_1}(\xi_1) \rangle \langle \psi_{m_1}(\xi_1)|\psi_{m_0}(\xi_0) \rangle |^2 ,
\nonumber \\
& &
\eea
which is independent of time.

We may now at some given finite time, $t_2$, change the parameters from $\xi_2 \rightarrow \xi_3$.
As before we find that for $t_2 \leq t<\infty$ the same probability as in eq. (7) is given by
\be
P_{n_0m_0}(t) = |\langle \psi_{n_0}(\xi_0)|\psi_{m_0}^{III}(t) \rangle |^2
\ee
where
\be
|\psi_{m_0}^{III}(t) \rangle = e^{-i H(\xi_3)(t-t_2)} |\psi_{m_0}^{II}(t_2) \rangle
\ee
The probability is now a function of time.

\section{Models}

In this chapter we consider systems that are topologically non-trivial, such as
one or two-dimensional topological insulators or topological superconductors.
The topological nature of these systems reveals itself both in the topological nature
of the groundstate of the infinite system and in the appearance of edge states if
the system is finite (bulk-edge correspondance). Different topological invariants
may be defined such as winding numbers for the one-dimensional examples considered
here and the Chern number for the two-dimensional superconductor considered later.
Both the winding numbers and the Chern number may be understood in various ways
\cite{hasan,alicea,yakovenko} and typically they count the number of edge modes
at the interface between the topological system and the vacuum.

Some examples are the models considered in this section which display both trivial
and topological phases. The dynamics of the edge modes of the topological
phases after a quantum quench is considered in sections 3-5.

\subsection{One-band spinless superconductor: the $1D$ Kitaev model}

The Kitaev one-dimensional superconductor with triplet p-wave pairing is described
by the Hamiltonian \cite{kitaev}
\bea
H &=& \sum_{j=1}^{\bar{N}} \left[ -{\tilde t} \left( c_j^{\dagger} c_{j+1} + c_{j+1}^{\dagger} c_j \right)
+ \Delta \left( c_j c_{j+1}+c_{j+1}^{\dagger} c_j^{\dagger} \right) \right] \nonumber \\
& - & \sum_{j=1}^N \mu \left( c_j^{\dagger} c_j -\frac{1}{2} \right)
\eea
where $\bar{N}=N$ if we use periodic boundary conditions (and $N+1=1$) or
$\bar{N}=N-1$ if we use open boundary conditions. Here $N$ is the number of sites. ${\tilde t}$ is the hopping amplitude
taken as the unit of energy, $\Delta$ is the pairing amplitude and $\mu$ the chemical potential. The operator
$c_j$ destroys a spinless fermion at site $j$.
 
In momentum space the model is written as
\begin{eqnarray}
\hat H = \frac 1 2\sum_k  \left( c_k^\dagger ,c_{-k}   \right)
H_k
\left( \begin{array}{c}
c_{k} \\  c_{-k}^\dagger  \end{array}
\right)
\end{eqnarray}
where
\begin{eqnarray}
H_k = \left(\begin{array}{cc}
\epsilon_k -\mu & i \Delta \sin k \\
-i \Delta \sin k & -\epsilon_k +\mu  \end{array}\right)
\nonumber \\
\end{eqnarray}
with $\epsilon_k=-2{\tilde t} \cos k$. Here $c_k$ is the Fourier transform of $c_j$.

\begin{figure}[t]
\begin{center}
\includegraphics[width=0.6\textwidth]{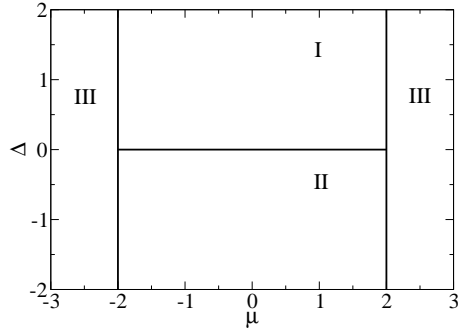}
\end{center}
\caption{\label{fig1}
Phase diagram of $1D$ Kitaev model. In phases I and II there are edge modes.
}
\end{figure}

In general, a  fermion operator may be writen in terms of two hermitian operators, $\gamma_1, \gamma_2$, in the
following way
\bea
c_{j,\sigma} &=& \frac{1}{2} \left( \gamma_{j, \sigma, 1} + i \gamma_{j, \sigma, 2} \right) \nonumber \\
c_{j,\sigma}^{\dagger} &=& \frac{1}{2} \left( \gamma_{j, \sigma, 1} - 
i \gamma_{j, \sigma, 2} \right) 
\eea
The index $\sigma$ represents internal degrees of freedom of the fermionic operator, such as spin
and/or sublattice index, the $\gamma$ operators are hermitian and satisfy a Clifford algebra
\be
\{\gamma_m,\gamma_n \}=2 \delta_{nm} .
\ee
In the case of the Kitaev model it is enough to consider $c_j=(\gamma_{j,1}+i \gamma_{j,2})/2$, since
the fermions are spinless.
In terms of these hermitian (Majorana) operators we may write that the Hamiltonian is given by,
using open boundary conditions,
\bea
H &=& \frac{i}{2} \sum_{j=1}^{N-1} \left[ 
(-{\tilde t}+\Delta) \gamma_{j,1} \gamma_{j+1,2}
+ ({\tilde t}+\Delta) \gamma_{j,2} \gamma_{j+1,1} \right] 
\nonumber \\
&-& \frac{i}{2} \sum_{j=1}^N \mu \gamma_{j,1}
\gamma_{j,2}
\eea

Taking $\mu=0$ and selecting the
special point ${\tilde t}=\Delta$ the Hamiltonian simplifies considerably to
\be
H(\mu=0, {\tilde t}=\Delta) = i {\tilde t} \sum_{j=1}^{N-1} \gamma_{j,2} \gamma_{j+1,1}
= -i {\tilde t} \sum_{j=1}^{N-1} \gamma_{j+1,1} \gamma_{j,2} 
\ee
Note that the operators $\gamma_{1,1}$ and $\gamma_{N,2}$ are missing
from the Hamiltonian. Therefore there are two zero energy modes. Defining from
these two Majorana fermions a single usual fermion operator (non-hermitian),
taking one of the Majorana operators as the real part and the other as the imaginary
part, its state may be either occupied or empty with no cost in energy.
Defining $d_j=1/2 \left( \gamma_{j,2}+i \gamma_{j+1,1} \right)$ and
$d_N=1/2 \left(\gamma_{N,2}+i \gamma_{1,1} \right)$ we can write the Hamiltonian as
\be
H={\tilde t} \sum_{j=1}^{N-1} \left( 2 d_j^{\dagger} d_j -1 \right) + \epsilon_N \left(2 d_N^{\dagger} d_N
-1 \right)
\ee
with $\epsilon_N=0$. Therefore the fermionic mode $d_N$ does not appear in the Hamiltonian
and the state may be occuppied or empty ($d_N^{\dagger} d_N=1,0$, respectively) with no energy
cost. These two states are therefore degenerate in energy
and are perfectly localized at the edges
of the chain as $\delta$-function peaks (with exponential accuracy as the system size grows).

The phase diagram of the Kitaev model shows 
three types of phases (see Fig. \ref{fig1}): two topological
phases in which there are gapless edge modes, if the system is finite, and two
trivial phases with no edge modes. In the various phases the bulk of the system is gapped
and at the transition lines the gap closes, allowing the possibility of a change
of topology. The transition lines are located at $\Delta=0, |\mu|\leq 2\tilde{t}$ and at 
$|\mu|=2 {\tilde t}$ and any $\Delta$.

\subsection{Multiband system: $1D$
Two-band Shockley model}

The Shockley model is a model of a dimerized system of spinless fermions with alternating
nearest-neighbor hoppings,  given by the Hamiltonian (see for instance \cite{yakovenko})
\be
H=\sum_{j=1}^N \psi^{\dagger}(j) \left[ U \psi(j) + V \psi(j-1) + V^{\dagger} \psi(j+1) 
\right)
\ee
where the $2 \times 2$ matrices $U$ and $V$ 
and the spinor $\psi$ representing two orbitals at site $j$ that are hybridized
by the matrices $U$ and $V$
are given by
\be
U= \left(\begin{array}{cc}
0 & t_1^* \\
t_1 & 0  \end{array}\right); 
V= \left(\begin{array}{cc}
0 & t_2^* \\
0 & 0  \end{array}\right);
\psi(j) =
\left(\begin{array}{c}
c_{j,A} \\
c_{j,B} \end{array}\right).
\ee
$t_1$ and $t_2$ are hoppings and $c_{j,A}$ ($c_{j,B}$) destroy spinless
fermions at site $j$ belonging to sublattice $A$ ($B$), respectively.

We may as well define Majorana operators as
\bea
c_{j,A} &=& \frac{1}{2} \left( \gamma_{j,A,1} + i \gamma_{j,A,2} \right)\nonumber \\
c_{j,B} &=& \frac{1}{2} \left( \gamma_{j,B,1} + i \gamma_{j,B,2} \right)
\eea
Here $A$ and $B$ take the role of pseudospins.
Taking $t_1$ and $t_2$ real, the Hamiltonian may be written as
\bea
H&=&\frac{i t_1}{2} \sum_{j=1}^N \left( \gamma_{j,A,1} \gamma_{j,B,2} +
\gamma_{j,B,1} \gamma_{j,A,2} \right)  \nonumber \\
&+&\frac{t_2}{4} \sum_{j=2}^N \left( \gamma_{j,A,1} \gamma_{j-1,B,1} +
\gamma_{j,A,2} \gamma_{j-1,B,2}  \right) \nonumber \\
&+& \frac{i t_2}{4} \sum_{j=2}^N \left( \gamma_{j,A,1} \gamma_{j-1,B,2}- 
i \gamma_{j,A,2} \gamma_{j-1,B,1}  \right) \nonumber \\
&+&\frac{t_2}{4} \sum_{j=1}^{N-1} \left( \gamma_{j,B,1} \gamma_{j+1,A,1} +
\gamma_{j,B,2} \gamma_{j+1,A,2} \right) \nonumber \\ 
&+& \frac{i t_2}{4} \sum_{j=1}^{N-1} \left(  \gamma_{j,B,1} \gamma_{j+1,A,2}- 
i \gamma_{j,B,2} \gamma_{j+1,A,1}  \right)
\eea

Choosing $t_1=0$ we find that the Majorana fermions $\gamma_{1,A,1}$, $\gamma_{1,A,2}$,
$\gamma_{N,B,1}$ and $\gamma_{N,B,2}$ do not contribute and are zero energy modes.
These decoupled zero-energy
modes are fermionic in nature, since the decoupled Majoranas
are located at the two end sites, $A$ and $B$, respectively.
This point is characteristic of the topological phase as long as
the bulk gap does not vanish.
In the trivial phase there are no decoupled Majorana operators.
As discussed for instance in Ref. \cite{yakovenko} the two types of
phases may also be distinguished by the winding number.

\subsection{Multiband system: $1D$ SSH model with triplet pairing}

\begin{figure}[t]
\begin{center}
\includegraphics[width=0.6\textwidth]{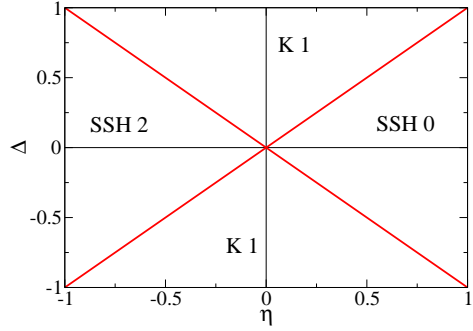}
\end{center}
\caption{\label{fig2}
(Color online)
Phase diagram of $1D$ SSH-Kitaev model for $\mu=0$. The phase SSH0 is trivial and
has no edge modes. In the phases K1 there is one edge Majorana mode at each edge
and in the phase SSH2 there are fermionic edge modes at each edge}
\end{figure}

This model may be viewed as a dimerized Kitaev superconductor \cite{tanaka}.
The dimerization is parametrized by $\eta$ and the superconductivity
by $\Delta$.

This model is given by the Hamiltonian
\bea
H = -\mu & \sum_j & \left(  c_{j,A}^{\dagger} c_{j,A} + c_{j,B}^{\dagger} c_{j,B} \right)
\nonumber \\
-{\tilde t} & \sum_j &  \left[  (1+\eta) c_{j,B}^{\dagger} c_{j,A} + (1+\eta) c_{j,A}^{\dagger} c_{j,B} \right.
\nonumber \\
&+& \left. (1-\eta) c_{j+1,A}^{\dagger} c_{j,B} +(1-\eta) c_{j,B}^{\dagger} c_{j+1,A} \right]
\nonumber \\
+\Delta  & \sum_j &  \left[  (1+\eta) c_{j,B}^{\dagger} c_{j,A}^{\dagger} + (1+\eta) c_{j,A} c_{j,B} \right.
\nonumber \\
&+& \left. (1-\eta) c_{j+1,A}^{\dagger} c_{j,B}^{\dagger} +(1-\eta) c_{j,B} c_{j+1,A} \right]
\nonumber \\
& &
\eea
(${\tilde t}$ is the hopping, $\Delta$ the pairing amplitude and $\mu$ the chemical potential).
The model with no superconductivity ($\Delta=0$) is related to the Shockley model
taking $t_1={\tilde t}(1+\eta)$ and $t_2={\tilde t}(1-\eta)$. The region of $\eta>0$ corresponds to $t_1>t_2$
and vice-versa for $\eta<0$.
The Hamiltonian in real space mixes nearest-neighbor sites and also has
local terms.

In terms of Majorana operators the Hamiltonian is written as
\bea
H &=& -\frac{\mu}{2} \sum_{j=1}^N \left( 2+i\gamma_{j,A,1} \gamma_{j,A,2}
+ i \gamma_{j,B,1} \gamma_{j,B,2} \right) \nonumber \\
&-& \frac{i{\tilde t}}{2} (1+\eta) \sum_{j=1}^N \left(
\gamma_{j,B,1} \gamma_{j,A,2} + \gamma_{j,A,1} \gamma_{j,B,2} \right) \nonumber \\
&-& \frac{i{\tilde t}}{2} (1-\eta) \sum_{j=1}^{N-1} \left(
\gamma_{j+1,A,1} \gamma_{j,B,2} + \gamma_{j,B,1} \gamma_{j+1,A,2} \right) \nonumber \\
&+& \frac{i\Delta}{2} (1+\eta) \sum_{j=1}^{N} \left(
\gamma_{j,A,1} \gamma_{j,B,2} + \gamma_{j,A,2} \gamma_{j,B,1} \right) \nonumber \\
&+& \frac{i\Delta}{2} (1-\eta) \sum_{j=1}^{N-1} \left(
\gamma_{j,B,1} \gamma_{j+1,A,2} + \gamma_{j,B,2} \gamma_{j+1,A,1} \right)
\nonumber \\
& &
\eea

Consider once again a vanishing chemical potential.
Taking $\eta=-1$ and
$\Delta=0$ we have a state similar to the SSH or Shockley models with two
fermionic-like zero energy edge states, since the four operators
$\gamma_{1,A,1}, \gamma_{1,A,2}; \gamma_{N,B,1}, \gamma_{N,B,2}$ are missing from
the Hamiltonian.
If we select $\eta=0$ and $t=\Delta$ is a Kitaev like state since there are two Majorana operators
missing from the Hamiltonian, $\gamma_{1,A,1}$ and $\gamma_{N,B,2}$, one from each end.
An example of a trivial phase is the point $\eta=1$ and $\Delta=0$ in which case there
are no zero energy edge states.
In Fig. \ref{fig2} the phase diagram is shown.
This model provides a testing ground for the comparison between fermionic and Majorana
edge modes. In addition, in some regimes it displays finite energy modes that are localized
at the edges of the chain, as obtained before in other multiband models \cite{rio}.

\subsection{Two-dimensional spinfull triplet superconductor}

\begin{figure}[t]
\begin{center}
\includegraphics[width=0.6\textwidth]{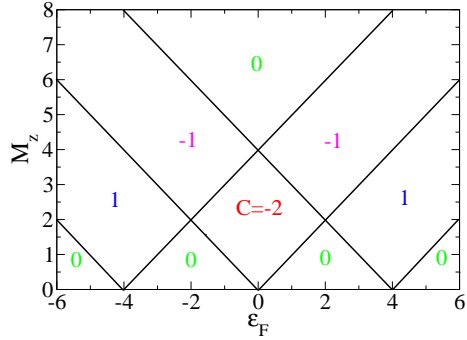}
\end{center}
\caption{\label{fig3}
(Color online)
Phase diagram of $2D$ p-wave model as a function of the chemical potential
and magnetization. $C$ is the Chern number of each phase associated with
the number of protected one-dimensional edge modes at finite magnetization.
}
\end{figure}

Another interesting case is that of a two-dimensional triplet superconductor
with $p$-wave symmetry, spin-orbit coupling and a Zeeman term \cite{sato}.
We write the Hamiltonian for the bulk system in momentum space as
\begin{eqnarray}
\hat H = \frac 1 2\sum_\vk  \left( {\boldsymbol \psi}_{\vk}^\dagger ,{\boldsymbol \psi}_{-\vk}   \right)
\left(\begin{array}{cc}
\hat H_0(\vk) & \hat \Delta(\vk) \\
\hat \Delta^{\dagger}(\vk) & -\hat H_0^T(-\vk) \end{array}\right)
\left( \begin{array}{c}
 {\boldsymbol \psi}_{\vk} \\  {\boldsymbol \psi}_{-\vk}^\dagger  \end{array}
\right)
\label{bdg1}
\end{eqnarray}
where $\left( {\boldsymbol \psi}_{\vk}^{\dagger}, {\boldsymbol \psi}_{-\vk} \right) =
\left( \psi_{\vk\ua}^{\dagger}, \psi_{\vk\da}^\dagger ,\psi_{-\vk\ua}, \psi_{-\vk\da}   \right)$
and
\begin{equation}
\hat H_0=\epsilon_\vk\sigma_0 -M_z\sigma_z + \hat H_R\,.
\end{equation}
Here, $\epsilon_{\boldsymbol{k}}=-2 \tilde{t} (\cos k_x + \cos k_y )-\veps_F$
is the kinetic part, $\tilde{t}$ denotes the hopping parameter set in
the following as the energy scale ($\tilde{t}=1$), 
$\boldsymbol{k}$ is a wave vector in the $xy$ plane, and we have taken
the lattice constant to be unity. Furthermore, $M_z$
is the Zeeman splitting term responsible for the magnetization,
in $\tilde{t}$ units.
The Rashba spin-orbit term is written as
\begin{equation}
\hat H_R = \boldsymbol{s} \cdot \boldsymbol{\sigma} = \alpha
\left( \sin k_y \sigma_x - \sin k_x \sigma_y \right)\,,
\end{equation}
 where
$\alpha$ is measured in the same units
 and $\boldsymbol{s} =\alpha(\sin k_y,-\sin k_x, 0)$.
The matrices $\sigma_x,\sigma_y,\sigma_z$ are
the Pauli matrices acting on the spin sector, and $\sigma_0$ is the
$2\times 2$ identity.
The pairing matrix reads
\begin{equation}
\hat \Delta = i\left( {\boldsymbol d}\cdot {\boldsymbol\sigma} \right) \sigma_y =
 \left(\begin{array}{cc}
-d_x+i d_y & d_z \\
d_z & d_x +i d_y
\end{array}\right)\,.
\end{equation}
We consider here $d_z=0$.
If the spin-orbit coupling is strong it is energetically favorable that
the pairing is of the form ${\boldsymbol d}=d \boldsymbol{s}$.

The energy eigenvalues and eigenfunction may be obtained solving the Bogoliubov-de Gennes equations
\be
\label{bdg2}
\left(\begin{array}{cc}
\hat H_0(\vk) & \hat \Delta(\vk) \\
\hat \Delta^{\dagger}(\vk) & -\hat H_0^T(-\vk) \end{array}\right)
\left(\begin{array}{c}
u_n\\
v_n
\end{array}\right)
= \epsilon_{\boldsymbol{k},n}
\left(\begin{array}{c}
u_n\\
v_n
\end{array}\right).
\ee
The 4-component spinor can be written as
\be
\left(\begin{array}{c}
u_n\\
v_n
\end{array}\right)=
\left(\begin{array}{c}
u_n(\boldsymbol{k},\uparrow) \\
u_n(\boldsymbol{k},\downarrow) \\
v_n(-\boldsymbol{k},\uparrow) \\
v_n(-\boldsymbol{k},\downarrow) \\
\end{array}\right) .\ee

The superconductor we consider here
is time-reversal invariant if the Zeeman term is absent.
The system then belongs to the symmetry class DIII where the topological invariant is
a $\mathbb{Z}_2$ index \cite{symmetry}.
If the Zeeman term is finite, time reversal symmetry (TRS) is broken and the system belongs
to the symmetry class D.
The topological invariant that characterizes this phase is the first Chern number $C$,
and the system is said to be a $\mathbb{Z}$~topological superconductor.
The phase diagram is shown in Fig. \ref{fig3}.

Due to the bulk-edge correspondence if the system is placed in a strip geometry and the system
is in a topologically non-trivial phase, there are robust edge states, in a number of pairs given by
the Chern number, if time reversal symmetry is broken. There are also counterpropagating edge states
in the $Z_2$ phases even though the Chern number vanishes, as in the spin Hall effect. In these
phases time reversal symmetry is preserved and the Kramers pairs of edge states give opposite
contributions to the Chern number. Interestingly, turning on the magnetization (Zeeman field)
time reversal symmetry is broken and the edge states are no longer topologically protected. However,
it was found that, even in regimes where $C=0$, there are edge states, reminiscent of the edge
states of the $Z_2$ phases.

\section{Dynamics of edge modes of $1D$ Kitaev model}

\subsection{Single quench}

\begin{figure}[t]
\begin{center}
\includegraphics[width=0.75\textwidth]{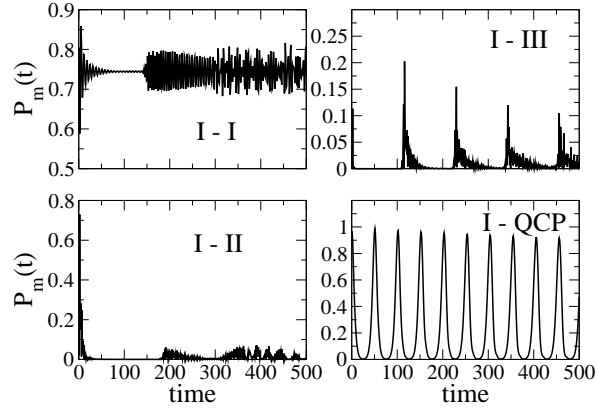}
\end{center}
\caption{\label{fig4}
Survival probability of the Majorana state of the one-dimensional
Kitaev model for different transitions across the phase diagram:
i) transition within
the same topological phase, $I$, ($\mu=0.5,\Delta=0.6) \rightarrow (\mu=1.0,\Delta=0.6)$,
ii) transition from the topological phase $I$ to the trivial phase $III$
($\mu=0.5,\Delta=0.6) \rightarrow (\mu=2.2,\Delta=0.6)$,
iii) transition from the topological phase $I$ with positive $\Delta$ to the topological phase $I$
with negative $\Delta$
($\mu=0.5,\Delta=0.6) \rightarrow (\mu=0.5,\Delta=-0.6)$,
iv) transition within
the same topological phase, $I$, to the quantum critical point ($\mu=0,\Delta=0.1) \rightarrow (\mu=0,\Delta=0)$
where the system is gapless.
The system has $100$ sites. Reproduced from Ref. \cite{ref1}.
}
\end{figure}

The stability of the Majorana fermions in this model has been considered recently
\cite{rajak}. 
In Fig. \ref{fig4} we present results for the survival probability of the
Majorana mode for different quenches \cite{ref1}. 
In the first panel we consider the case of a quench within the same topological phase
clearly showing that the survival probability is finite.
Since the parameters change, there is a decrease of $P(t)$ as
a function of time due to the overlap with {\it all} the eigenstates of the chain with the new
set of parameters, but after some oscillations the survival rate stabilizes at some finite
value. As time grows, oscillations appear again centered around some finite value. 
Therefore the Majorana mode is robust to the quench.
In the second panel we consider a quench from the topological phase $I$ to the trivial, non-topological
phase $III$. The behavior is quite different. After the quench the survival probability decays fast to nearly zero.
After some time it increases sharply and repeats the decay and revival process. 
Similar results are found for a quench between the two topological phases $I$ and $II$.
As discussed in ref. \cite{rajak} the revival time scales with the system size. 
At this instant the wave function is peaked around the center of the system and is the result of
a propagating mode across the system with a given velocity and, therefore, scales with the
system size. In the infinite system limit the revival time will diverge and the Majorana mode
decays and is destroyed. A qualitatively different case is illustrated in the last panel of
Fig. \ref{fig2} where a quench from the topological phase $I$ to the quantum critical point at the
origin is considered. 

Let us analyse these oscillations in greater detail.
Consider first $\mu=0$ and quenches where one varies $\Delta$, or a fixed $\Delta$ and changing $\mu$.
In the case of $\mu=0$ the critical point is located at $\mu=0,\Delta=0$ and in the second case there
is a line of critical points at $\mu=2{\tilde t}$.
One finds that there is a point that separates the existence or not of oscillations.
If the initial state is close enough to the critical point there are
oscillations. Otherwise they are absent.
For instance, in the quench from the topological phase
$I$ to the critical point at $\mu=0, \Delta=0$, the point is located as
$N=100, \Delta=0.34$, $N=200, \Delta=0.18$, $N=400, 0.05<\Delta <0.1$.
In the vicinity of the two critical lines of points (around $\mu=2{\tilde t},\Delta=0$),
no matter how close the initial
point is to the critical line, one does not find oscillations (for further details see ref. \cite{ref3}).

\begin{figure}[t]
\begin{center}
\includegraphics[width=0.48\textwidth]{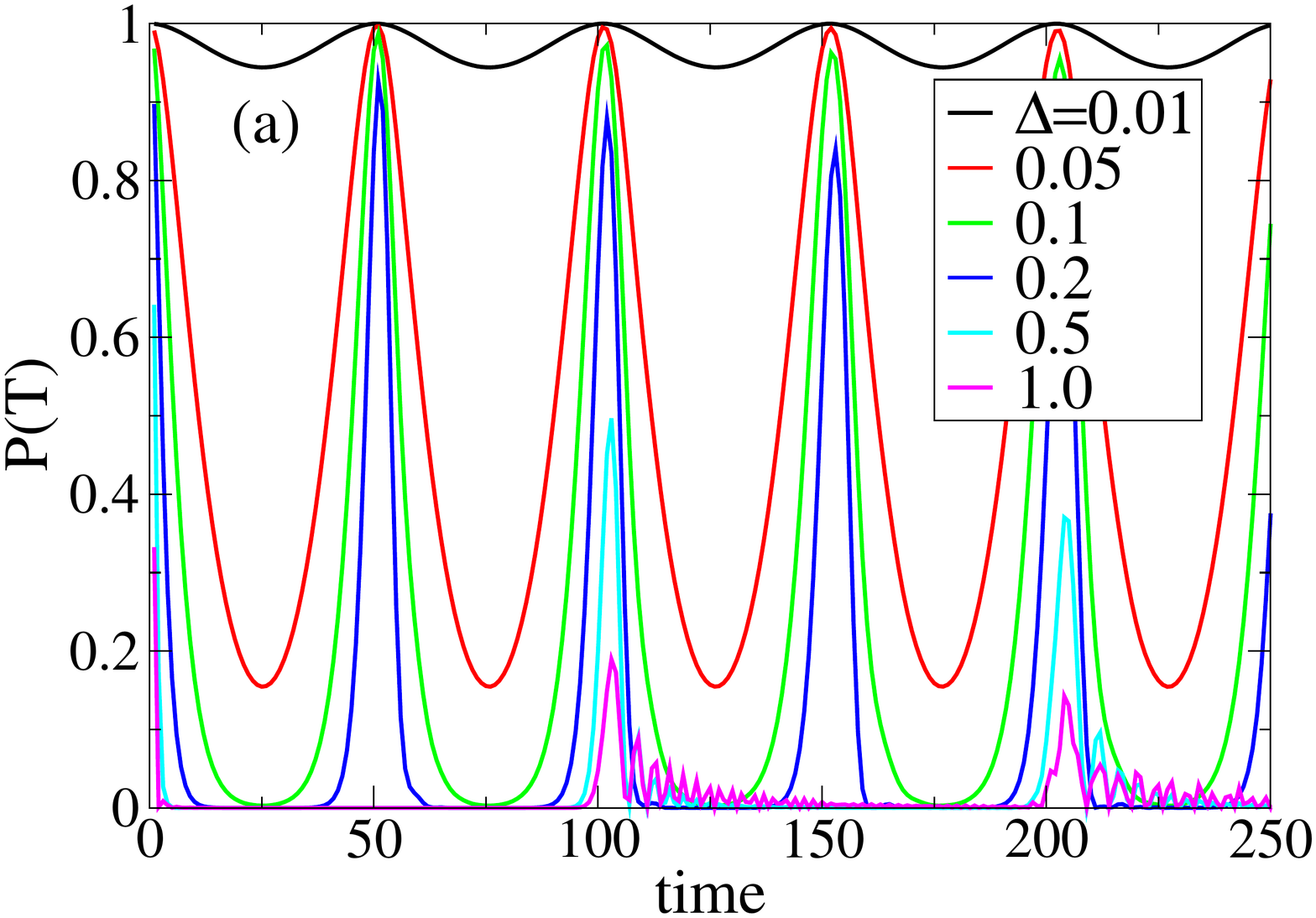}
\includegraphics[width=0.48\textwidth]{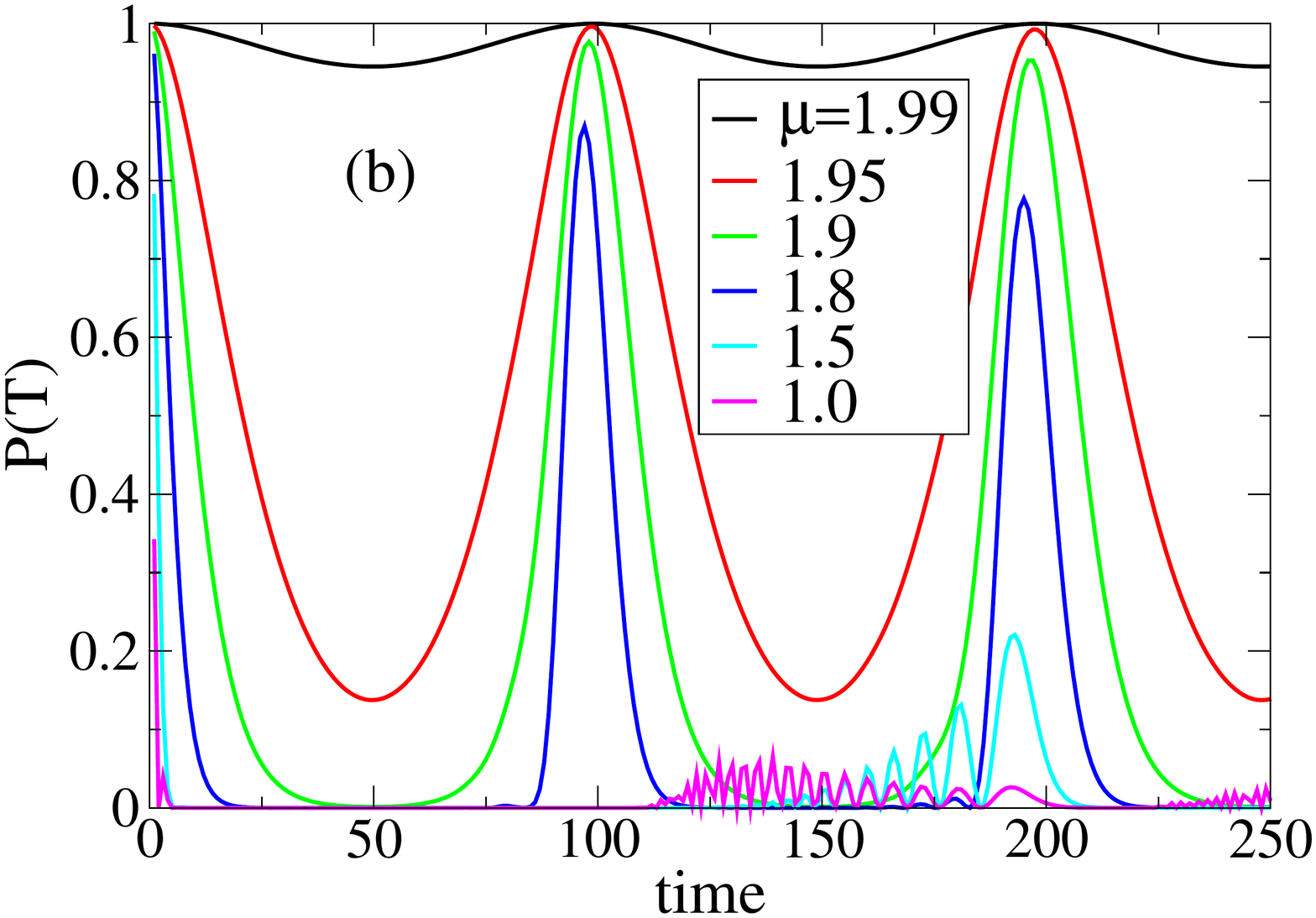}
\includegraphics[width=0.48\textwidth]{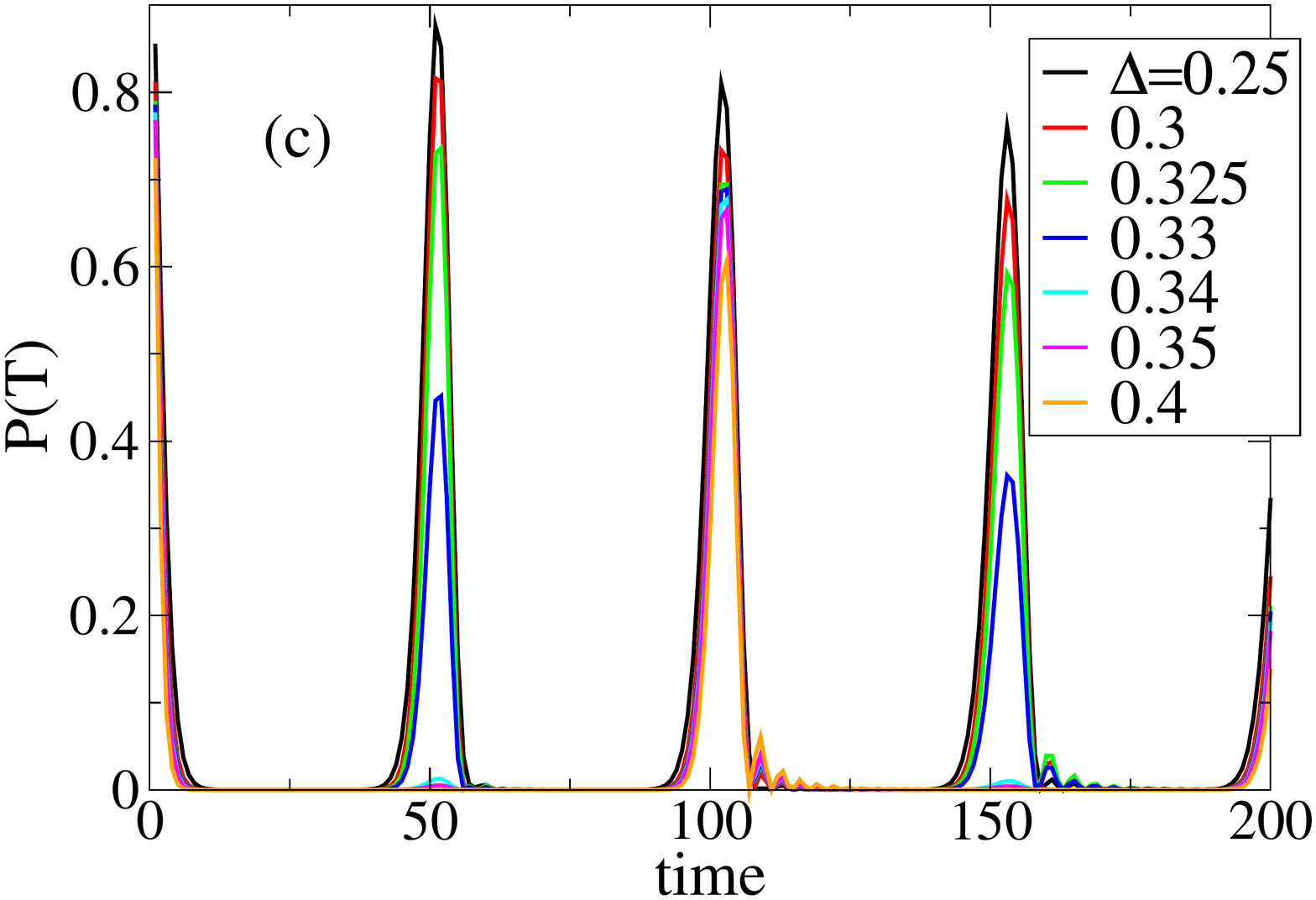}
\end{center}\caption{\label{fig5}
(Color online)
Survival probability, $P(t)$, of a Majorana mode in the $1D$ Kitaev model as one approaches
critical points. In panel (c) the crossover to period doubling is shown as one
approaches the critical region. The system size is $N=100$. In (a) $\mu=0$, in (b) $\Delta=0.5$,
and in (c) $\mu=0$. Reproduced from Ref. \cite{ref3}.
}\end{figure}

In Fig. \ref{fig5}
the survival probability, $P(t)$, of a Majorana mode as a function of time, for various critical quenches is presented.
In the first panel are shown the oscillations of $P(t)$ as one quenches from a given
value of $\Delta$ to the critical point $\mu=0, \Delta =0$, maintaining
$\mu=0$. 
For small deviations of the initial value of $\Delta$ from
the critical point, $P(t)$ is close to $1$ and as one increases the
distance from the critical point the amplitude decreases considerably.
The oscillations are quite smooth and clear until the amplitude has
decreased enough to reach zero. Beyond this point there is a periodicity
but no longer oscillations since there are increasing regions where $P(t)$
basically vanishes. In this case it seems more like the revival times
of non-critical quenches, even though the curves are still smooth. 
Beyond a given value of $\Delta_d$ there is a {\it period doubling}.
Also, after this period doubling the survival probability looses its regular
periodic behavior and shows more oscillations of smaller periods and amplitude
decays that are similar to results previously found in quenches away from
critical points \cite{rajak,ref1}. 
In the second panel are shown quenches to the critical line $\mu=2{\tilde t}$ keeping
$\Delta=0.5$ and decreasing the chemical potential. The behavior is similar
to the first panel.
In the third panel is shown in greater detail the {\it crossover} to period
doubling for the transition to the critical point. 
The point of crossover, $\Delta_d$, scales linearly with $1/N$.

\begin{figure}[t]
\begin{center}
\includegraphics[width=0.6\textwidth]{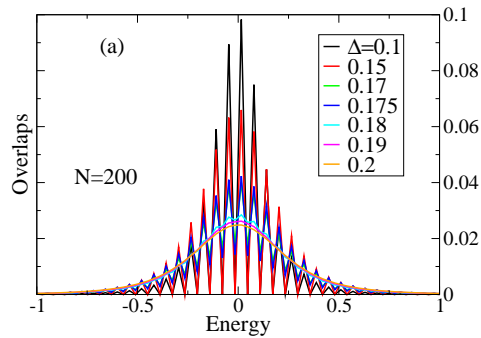}
\end{center}
\caption{\label{fig6}
(Color online)
Overlaps for the $1D$ Kitaev model as a function of energy. The critical point
is $\mu=0,\Delta=0$. Reproduced from Ref. \cite{ref3}.
}
\end{figure}

The survival probability is determined by the various energies of
the final Hamiltonian eigenstates and their overlaps to the initial single-particle state.
In Fig. \ref{fig6} the overlaps between the initial lowest
energy state (Majorana mode) and all the final state eigenvectors are shown,
as a function of their energies, for $N=200$. 
In general, the overlaps are peaked at the lowest energies. 
There is a clear separation of regimes as one reaches
the crossover region where the period doubling occurs. At small values
of $\Delta$ the overlaps oscillate between finite and zero
values. This is
a parity effect distinguishing even and odd number of sites. 
It can be noted that the overlaps are very sharp around the lowest
energy states. As the crossover occurs the overlaps are no longer zero
at some energy eigenvalues and actually become very smooth. This means
that the contributions from the various energy states changes, the
time behavior is affected and the clean oscillations are no longer
observed. In order to have clean oscillations one needs contributions
from few energy levels. A perfect oscillation requires finite overlaps
to two states and the frequency of the oscillations is the difference
in their energy values. In general, the overlaps have very different
magnitudes to the two states and the period of oscillations shown in
$P(t)$ depends on their magnitudes. Adding significant contributions
from other energy eigenstates leads first to modulated oscillations
and then to a complicated time dependence.

\begin{figure}[t]
\begin{center}
\includegraphics[width=0.48\textwidth]{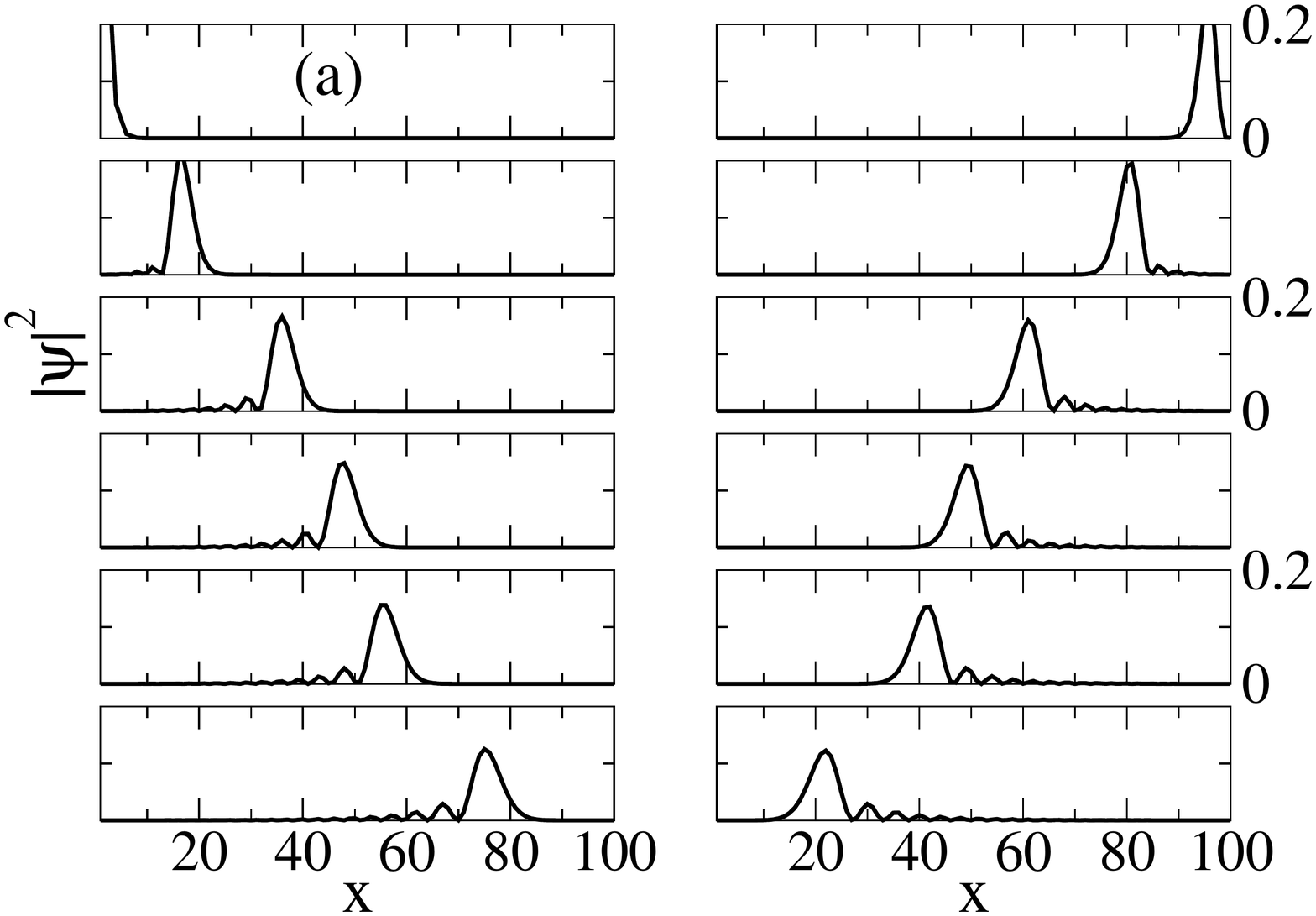}
\includegraphics[width=0.48\textwidth]{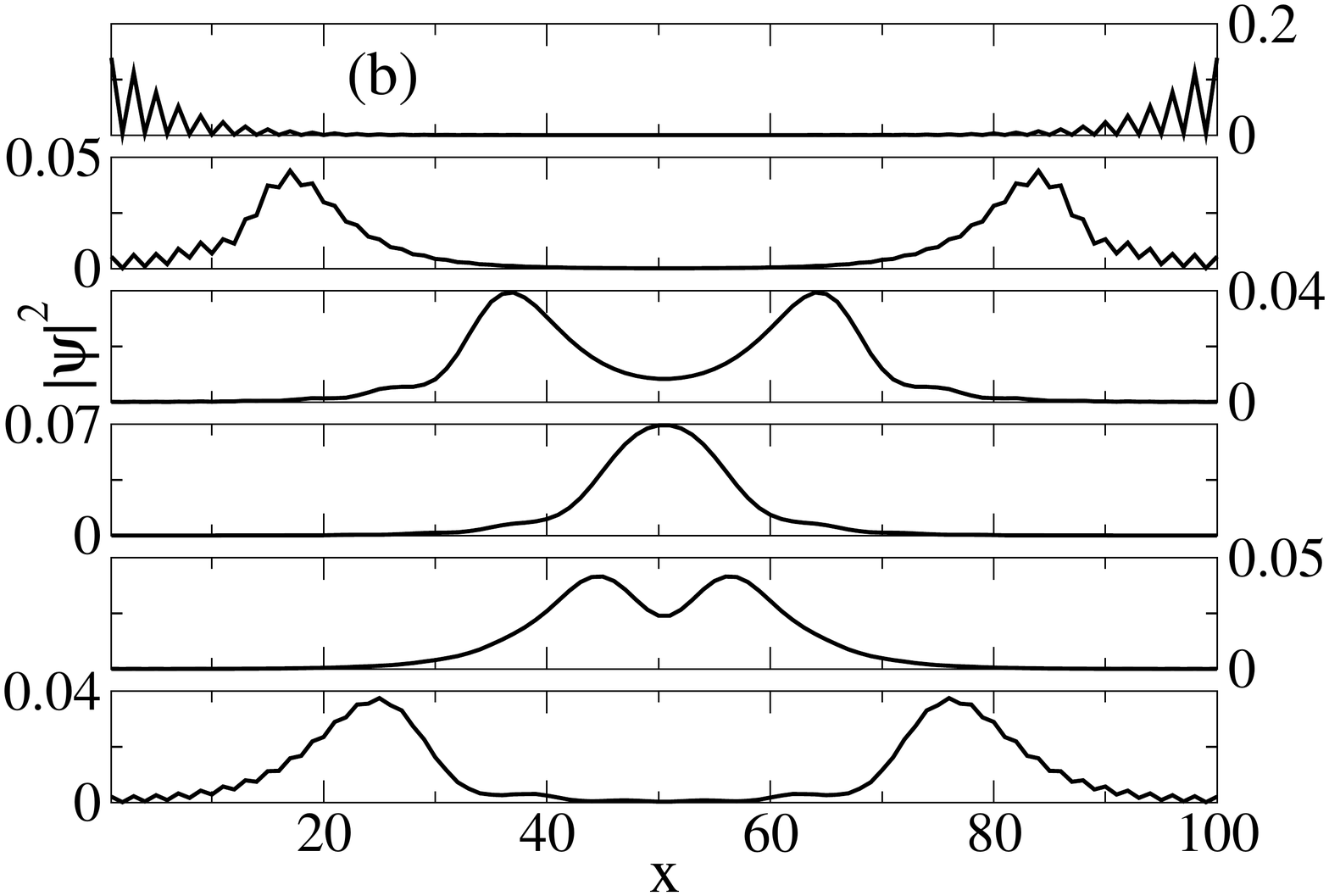}
\end{center}
\caption{\label{fig7}
Solitonic-like vs. constructive interference behavior of the wave functions in the $1D$ Kitaev model.
In (a) the initial state is far from the critical point (CP)
and in (b)one is close to the CP. In (a) the quench takes place from $\mu=0,\Delta=0.5$
to $\mu=0,\Delta=0$ and in (b) from $\mu=0,\Delta=0.1$ to the same CP. The results
are for a system size $N=100$. In (a) the sequence of times is $t=2,10,20,26,30,40$ and
in (b) $t=2,10,20,25,30,40$. Reproduced from Ref. \cite{ref3}.
}
\end{figure}

The origin of the period doubling is understood in the following way.
In Fig. \ref{fig7} the time evolution of the Majorana state is
shown, for a critical quench 
from a region far from the critical point where the period has doubled, and a quench 
from the region of oscillations, close to the critical point.
In the first
case the wave functions at each edge are separated in two energy modes
while for the second they are mixed. This is due to the long range
correlations close to the critical point that effectively decrease the
system size and lead to the coupling of the two edge modes. In the first
case the time evolved states from each edge cross each other in a solitonic
like behavior while in the second case there is a constructive interference
when the peaks of the evolved state meet at the center of the wire.
Consistently with the results for the overlaps, in this regime the difference
in energy
between states with high weight halves, and the period doubles.

Both the revival time and the period of the oscillations 
are associated with the propagation of the state along the system, with a velocity that, 
in the case of a free system, travels at a velocity given by the quasiparticle 
energy slope \cite{hamma}. 
In the case of interacting systems, it generalizes to a limiting velocity value, 
similar to a light-cone propagation \cite{lieb,bravyi,kuhr,bonnes}.

\begin{figure}[t]
\begin{center}
\includegraphics[width=0.6\textwidth]{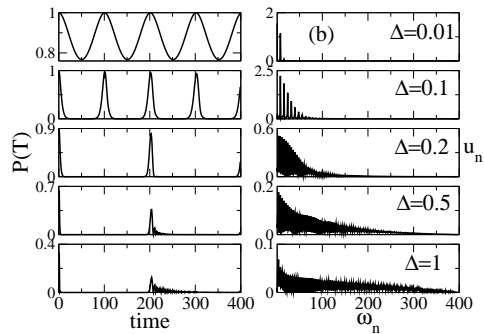}
\end{center}
\caption{\label{fig8}
Survival probability and its Fourier analysis for the $1D$ Kitaev model.
Reproduced from Ref. \cite{ref3}.
}
\end{figure}

A similar conclusion is obtained performing a Fourier
analysis of the time evolution of the survival probability. This is shown in Fig. \ref{fig8}.
While for small initial values of $\Delta$ the distribution is quite narrow
around low frequencies,
it changes significanly as $\Delta$ grows, becoming quite extended.
In the Fourier decomposition the amplitudes, $u_n$, are for the frequencies with
values $\omega_n=\pi (n-1)/N_t$, where $N_t$ is the number of time points considered.

It is also interesting to study the survival probability of excited states,
that in this problem are
extended states throughout the chain. 
Close to the critical point the survival probability of most states is close to $1$ except
near the low energy modes.  Further away from the critical point the deviation of the
survival probability from unity extends to higher energy states due to the orthogonality between the eigenstates
of the original and final Hamiltonians \cite{ref3}.

\subsection{Generation of Majorana states}

While quenches in general destabilize the edge states,
due to the finiteness of a system, we may generate Majorana states through a sudden quench
starting from a
trivial phase. Even though in the thermodynamic limit the topological
properties can not be changed by a unitary transformation \cite{ref1,rigol}, 
the probability that a given initial state in a trivial phase $III$
may collapse to a Majorana of the final state Hamiltonian in phase $I$ is finite
and independent of time.
Quenching to a state close to the transition line, the overlaps of several (extended)
states are considerable due to the spatial extent of the Majorana states. If the quench is
deeper into the topological phase these become more localized and the overlap decreases.
Interestingly the larger overlap is found for some higher energy, extended states.

A sequence of quenches allows for the manipulation of the states \cite{ref3}.
A possibility to turn off and on Majoranas can be trivialy seen in the following way.
Consider starting from a state inside region $I$ of the phase diagram Fig. \ref{fig1}. Perform a critical quench to the
line $\Delta=0$ and then a quench back to the original state. Choosing appropriately
$t_1$ we may get a state with no overlap with the initial Majoranas, as 
illustrated in Fig. \ref{fig5}.
So we are back to a topological phase but with no edge states. But Majoranas may
be switched back on if at a time $t_2>t_1$ we perform another quench to a state in
region $I$. 
Due to the
quench to $\xi_3$ a finite probability to find the Majorana state is found \cite{ref3} even
though if no quench from $\xi_2=\xi_0 \rightarrow \xi_3$ was performed, and having
chosen appropriately $t_1$, the survival probability of the Majorana states was
tuned to vanish.
Note that the overlap of Majorana state of $H(\xi_3)$ with a Majorana state of
$H(\xi_0)$ is finite, since the states are chosen to be close by.

\section{Dynamics of $1D$ multiband systems}

While in the previous section Majorana edge states of Kitaev's model were considered, edge states
in other systems, including topological insulators, have also been considered and
show similar properties. 
In this section we consider two topological systems, the Shockley model \cite{yakovenko}
which has fermionic edge states and no Majoranas, and the SSH-Kitaev model \cite{tanaka}
which displays both types of edge states in different parts of the phase diagram,
allowing a comparison of different edge state dynamics.

\begin{figure}[t]
\begin{center}
\includegraphics[width=0.75\textwidth]{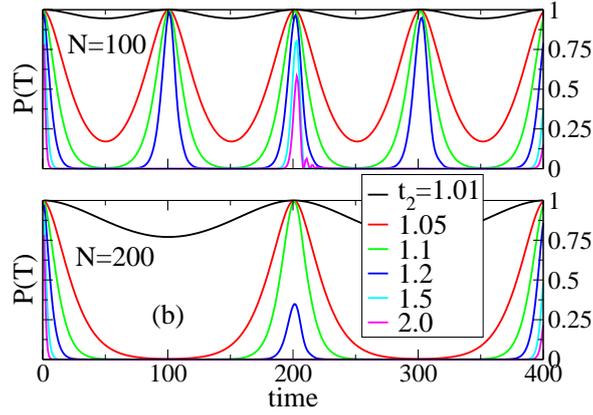}
\end{center}
\caption{\label{fig9}(Color online)
Survival probability of edge modes of the $1D$ Schockley model.
Critical quenches are considered from the
topological region to the transition point ($t_1=t_2$).
Reproduced from Ref. \cite{ref3}.
}
\end{figure}

\subsection{$1D$ Shockley model}

In Fig. \ref{fig9} we show critical quenches to a final state with
$t_2=t_1=1$ starting from different initial points in the topological
region ($t_2>t_1$). The cases of $N=100$ and $N=200$ are shown. As
in the Kitaev model the period scales with the system size. The behavior
is very similar to the Kitaev model. We see the period doubling for
both cases for $t_2=1.5$.
For $t_2=2.0$ the smoothness of the oscillations is replaced by a superposition
of many frequencies.
From the point of view of edge state dynamics the behavior of Majoranas and fermionic
edge states are similar.

Also, moving further away from the critical point a behavior similar to
Fig. \ref{fig4}b is seen with a rapid decrease of $P(t)$ and the appearance
of revival times.

\begin{figure}[t]
\begin{center}
\includegraphics[width=0.48\textwidth]{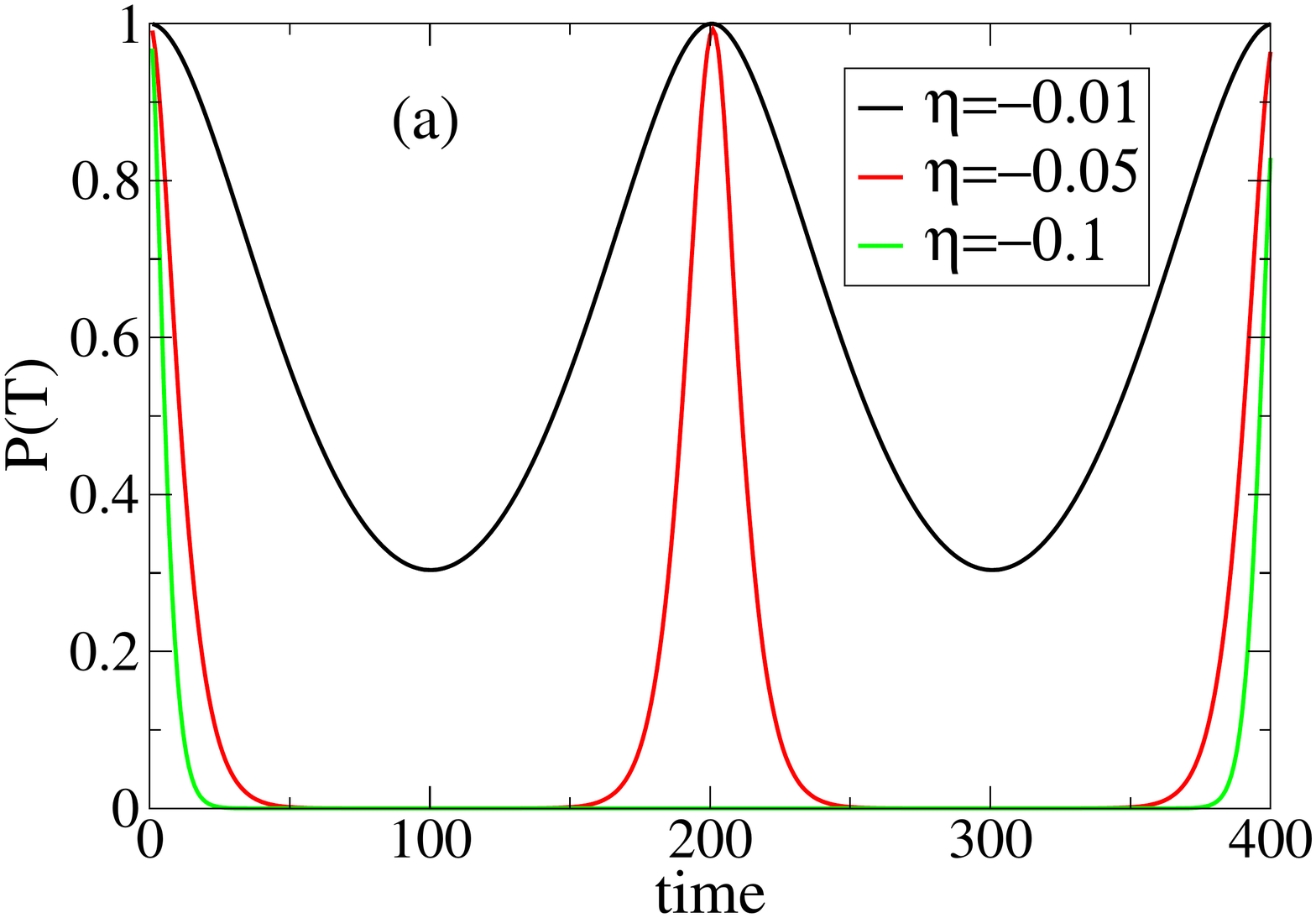}
\includegraphics[width=0.48\textwidth]{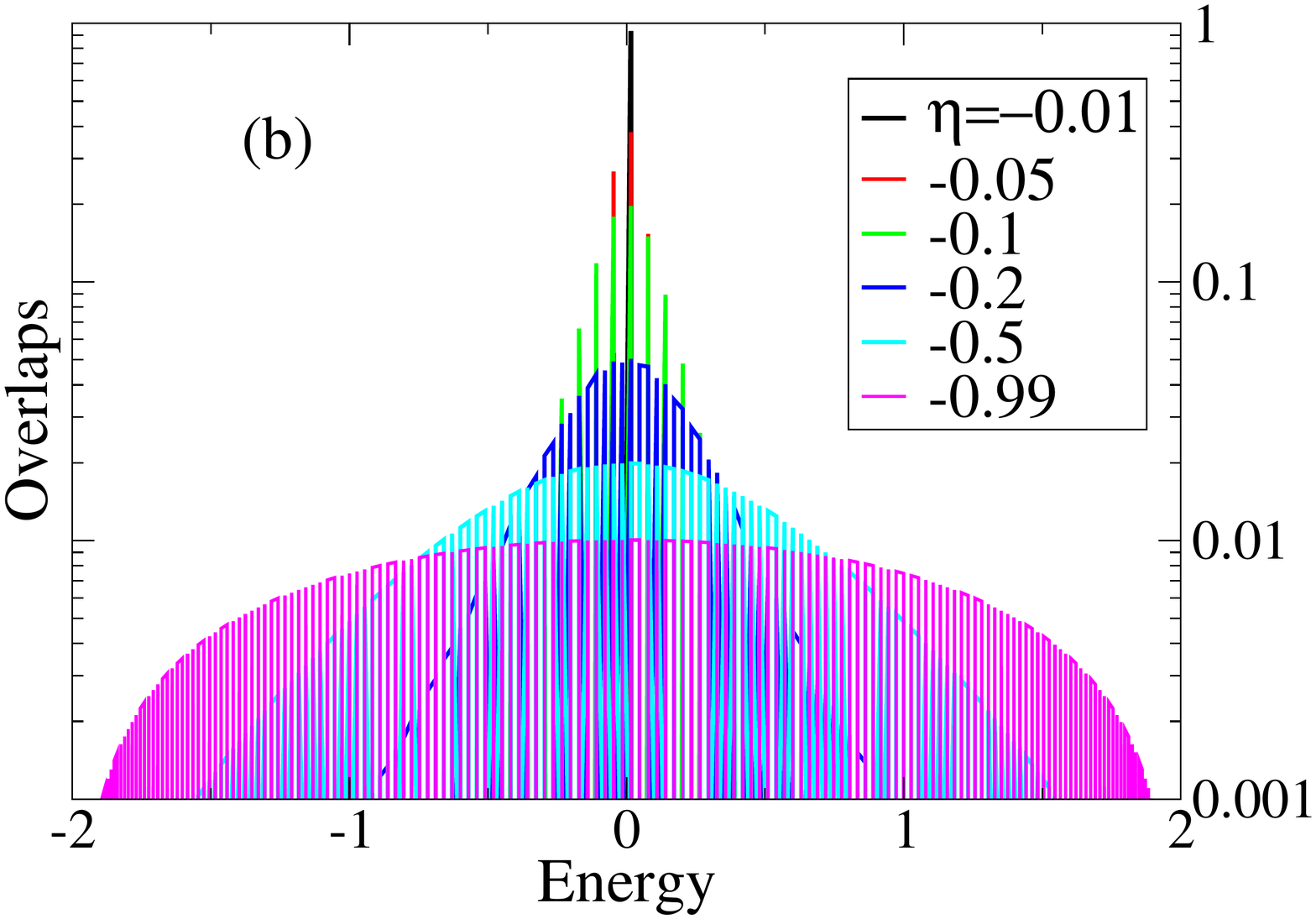}
\includegraphics[width=0.48\textwidth]{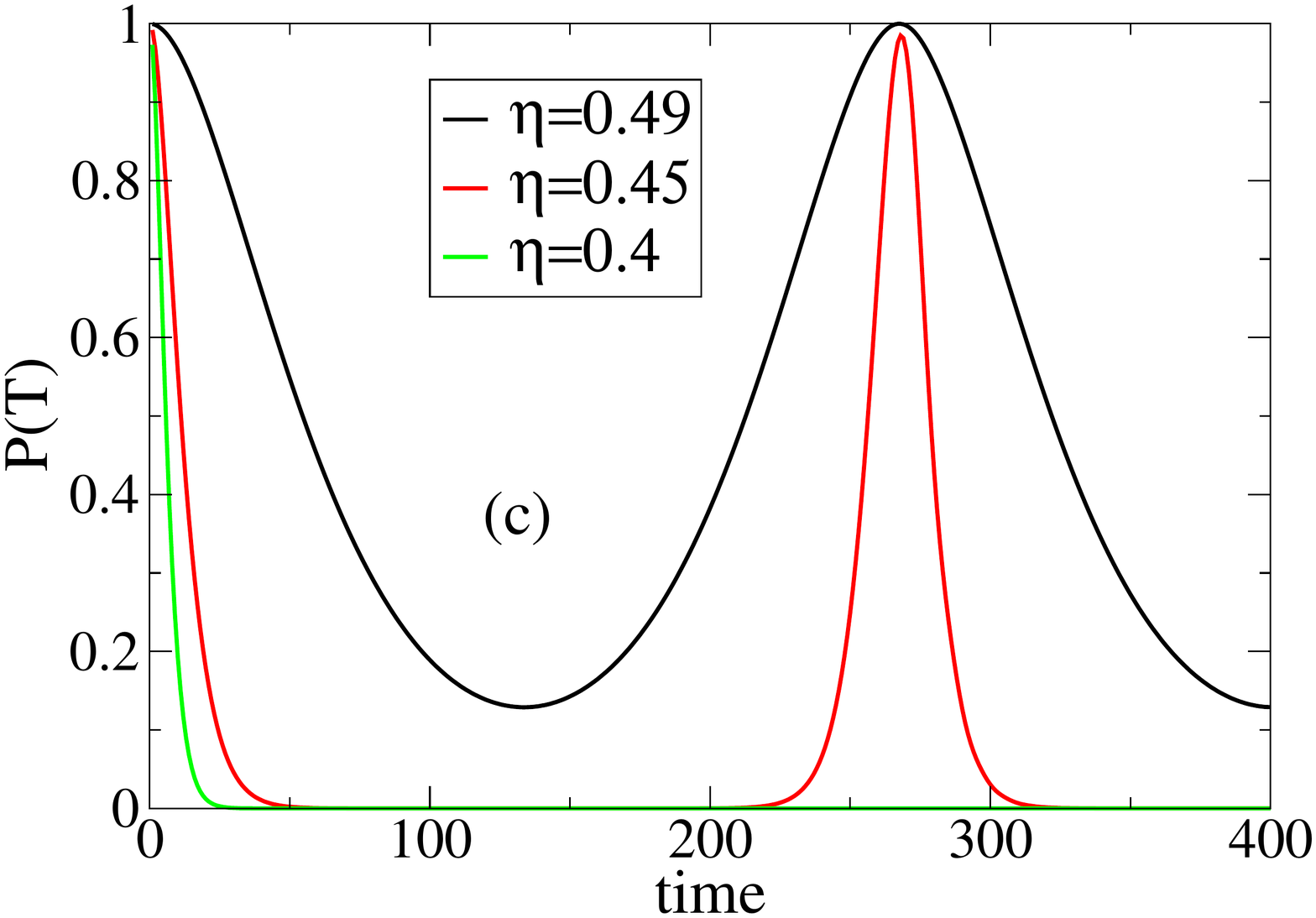}
\includegraphics[width=0.48\textwidth]{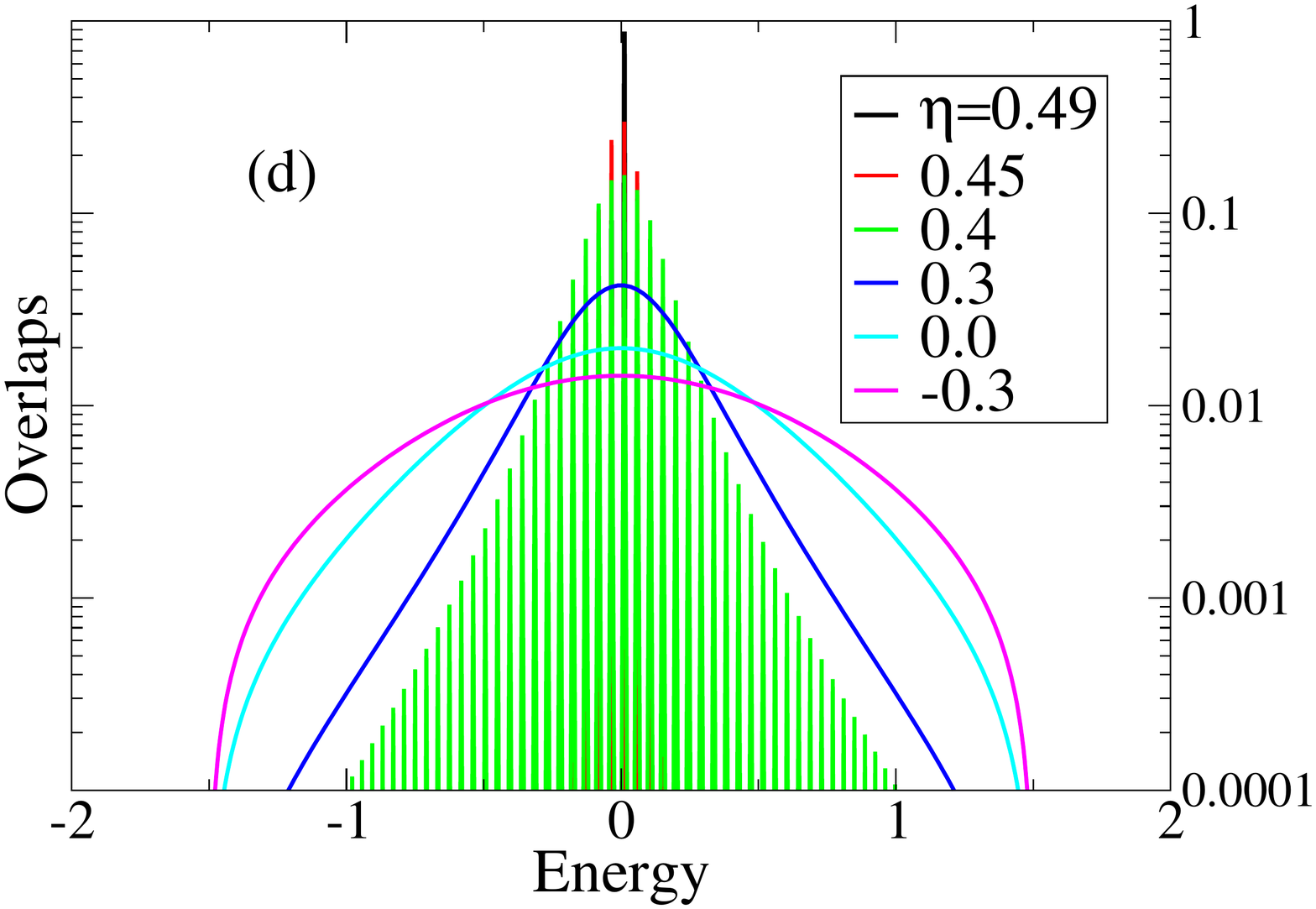}
\end{center}
\caption{\label{fig10}
(Color online)
Critical quenches in the $1D$ SSH-Kitaev model: survival probability and overlaps.
The CP on the first two panels is $\eta=0,\Delta=0$ (SSH2) and on the last two panels the CP is
$\eta=0.5, \Delta=0.5$ (K1).
Reproduced from Ref. \cite{ref3}.
}
\end{figure}

\subsection{$1D$ SSH-Kitaev model}

The similarities between Majorana and fermionic edge states are further shown considering the SSH-Kitaev model.
In Fig. \ref{fig2} we showed the phase diagram of the SSH-Kitaev model \cite{tanaka}
in the case of $\mu=0$. 
In phase K1 we are in the Kitaev regime with one zero energy edge mode at each edge
(Majoranas). In the SSH regimes we are closer to the behavior of the SSH model with
fermionic modes. In SSH 0 there are no edge modes. In SSH 2 there are two zero energy
fermionic modes.

In Fig. \ref{fig10} 
we consider critical quenches to points in the transition between different
topological regions. In the top panels we consider $P(t)$ and the overlaps, respectively, of a transition
at $\mu=0$ from the SSH 2 regime to the critical point $\eta=0, \Delta=0$ by considering
different initial values of $\eta=-0.01,-0.05,-0.1,-0.2,-0.5,-0.99$. In the lower
panels we consider critical quenches to the critical point $\eta=0.5, \Delta=0.5$ changing
the initial value of $\eta$. In both cases note that there is again a change of the distribution
of the overlaps from sharp peaks, at small deviations from the critical point,
to a broad distribution of the overlaps as one moves sufficiently away from the
critical point; again there is a crossover between the two regimes (not shown), 
as for the Kitaev model. However, the overlaps are not smooth as a function of energy.
Note that in the first case $\Delta=0$, which means that this occurs in the context of the
SSH model with no superconductivity. In the second case we have a mixture of SSH and
Kitaev model, but the behavior is qualitatively similar in the crossover region. Beyond
it we find again the very smooth distributions of the overlaps as in the Kitaev model.

\section{Dynamics of edge states of 2D triplet superconductor}

\subsection{Wave-function propagation}

\begin{figure}[t]
\begin{center}
\includegraphics[width=0.32\textwidth]{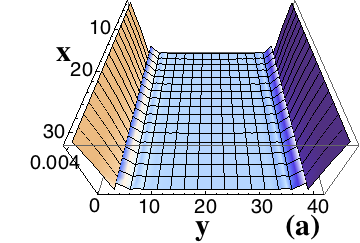}
\includegraphics[width=0.32\textwidth]{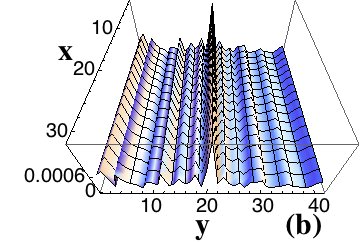}
\includegraphics[width=0.32\textwidth]{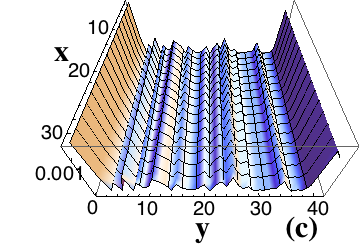}
\end{center}
\caption{\label{fig11}
(Color online)
Time evolution of real space $|u_{\uparrow}|^2$ for $(M_z=2,\mu=-5) \rightarrow (M_z=0,\mu=-5)$ 
$C=1 \rightarrow C=0$ (trivial)
for $t=0,t=50,t=62$, shown in (a), (b), (c), respectively.
Note that in these transitions there are no edge states in the final
states. In the various panels the horizontal axis is the $y$ direction and the vertical direction is the $x$ direction.
The system size is $31 \times 41$.
Reproduced from Ref. \cite{ref1}.
}
\end{figure}

The edge states appear if we consider a strip geometry of finite transversal
width, $N_y$, with open boundary conditions (OBC) along $y$
and periodic boundary conditions (PBC) along the longitudinal direction, $x$, of size $N_x$.
The diagonalization of this Hamiltonian expressed in real space involves the solution of a 
$(4 N_x N_y) \times (4 N_x N_y)$ eigenvalue problem.
The energy states include states in the bulk and states along the edges and are
written in the form of a 4-component spinor as
\be
\left(\begin{array}{c}
u_n\\
v_n
\end{array}\right)=
\left(\begin{array}{c}
u_n(j_x,j_y,\uparrow) \\
u_n(j_x,j_y,\downarrow) \\
v_n(j_x,j_y,\uparrow) \\
v_n(j_x,j_y,\downarrow) \\
\end{array}\right) 
\label{spinorrs}.\ee
Here $j_x,j_y$ are the spatial lattice coordinates along $x$ and $y$, respectively.
Focusing our attention on a Majorana mode, we present in Fig. \ref{fig11}
the time evolution of the absolute value of the spinor component 
$u_n(j_x,j_y,\uparrow)$, as an example,
for a time evolution for $(M_z=2,\mu=-5) \rightarrow (M_z=0,\mu=-5)$ 
$C=1 \rightarrow C=0$ (trivial)
for $t=0,t=50,t=62$, shown in (a), (b), (c), respectively.
The other spinor components have a qualitatively similar behavior. 
A set of characteristic time values are selected (time is expressed in units of $1/\tilde{t}$).
The initial state shows a mode that is very much peaked at the borders
of the system and that decays fast inside the supercondutor along the transverse
direction. 
As time evolves the peaks move towards the center until they merge at some later time, dependent
of the system transverse size (as for the Kitaev model). 
After this time the peaks move back from the center, the wave functions become more extended as a mixture
to all the eigenstates becomes more noticeable. 
Eventually at later times the wave function recovers
a shape that is close to the initial state and there is a partial revival of the original state. The process
then repeats itself but the same degree of coherence is somewhat lost. In Fig. \ref{fig11}
the quenches are carried out between a topological phase and a trivial phase ($C=1\rightarrow C=0$) and
($C=-2 \rightarrow C=0$). 
The behavior is therefore qualitatively the same as for the $1D$ case.

\begin{figure}[t]
\begin{center}
\includegraphics[width=0.6\textwidth]{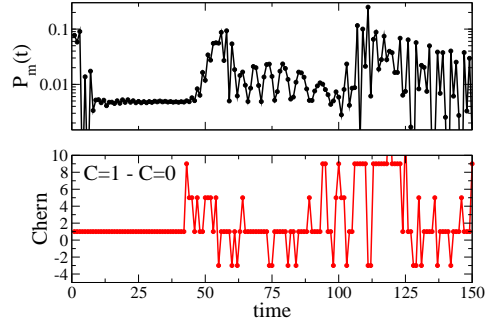}
\end{center}
\caption{\label{fig12}
(Color online)
Comparison of time evolution of (a) survival probability and (b) Chern number for the
case of strong spin orbit coupling for $(M_z=2,\mu=-5)\rightarrow (M_z=0,
\mu=-5)$ corresponding to $C=1\rightarrow C=0$.
The Chern number remains stable at the initial state value until
the Majorana mode reaches the middle point of the system. Beyond this
instant the Chern number fluctuates.
Reproduced from Ref. \cite{ref1}.
}
\end{figure}

\subsection{Evolution of Chern numbers}

The topology of each phase may be characterized by the Chern number, defined
over the Brillouin zone of the system \cite{sato}.
As the system evolves in time, the wave functions change. Solving for the evolution
of the wave functions we may calculate the Chern number as a function of time and
determine how the topology changes as well. Due to the fluctuating evolution
of the overlaps between a given state and all the others in the appropriate
subspace, we may expect that wave functions over the Brillouin zone will fluctuate
considerably as time goes by.

In Fig. \ref{fig12} it is shown that the Chern number remains locked to the initial
state value until the Majorana mode reaches the center point of the system and, therefore,
the topology is maintained. Beyond that instant
the Chern number starts to fluctuate which indicates that gaps are closing and opening
due to the time evolution. In the thermodynamic limit the revival times extend to
infinity and the Chern number does not change \cite{ref1,rigol}, even though the edge states
do decay. However, the Chern number may change due to the finiteness of the
system.
The values taken by the Chern number at a given time can be quite large. 
Since the Chern numbers fluctuate considerably it may make sense to look at the time
averaged Chern numbers. These average values have a very slow convergence to 
the value corresponding to the Chern value of the final state
and is not conclusive if it fully occurs.

\section{Periodic driving}

A different type of time perturbation that has attracted considerable interest are
periodic perturbations. 
While quenches, either abrupt or slow, in general destabilize the edge states, topological phases
can be induced by periodically driving the Hamiltonian
of a non-topological system, such as shown before in topological insulators \cite{oka,floquet0,lindner}
and in topological superconductors, with the appearance of Majorana fermions \cite{jiang,luo,xiaosen,viola}.
Their appearance in a one-dimensional p-wave superconductor was studied in Ref. \cite{liu}
and in Ref. \cite{manisha} introducing external periodic perturbations; the case of
intrinsic periodic modulation was also considered \cite{foster}.
The periodic driving leads to new topological states \cite{lindner}, and to a
generalization of the bulk-edge correspondence, that reveals a richer structure \cite{prx,balseiro} as
compared with the equilibrium situation \cite{symmetry,delplace}.
Similarly, in topological superconductors
new phases may be induced and manipulated due to the presence of the
periodic driving \cite{platero,hexagonal,liu}, such as shining a laser on
a topologically trivial system.

\subsection{Floquet formalism}

The time evolution of a state under the influence of a time dependent
Hamiltonian is given by
\be
i  \frac{\partial}{\partial t} \psi(k,t)=H(k,t) \psi(k,t)
\ee
where $k$ is the momentum, $t$ the time and we take $\hbar=1$.
We can decompose the Hamiltonian in two terms: a time independent one,
$H(k)$,
and an extra term due to the external time-dependent perturbation, that
we want to take as periodic with a given frequency, $\omega$,
\be
H(k,t)=H(k)+f(\omega t) H_d(k)
\ee
where $f(\omega t + 2\pi)=f(\omega t)$.
Here $H_d(k)$ is of the form of the unperturbed Hamiltonian but with
only one non-vanishing term.
Looking for a solution of the type 
\be
\psi(k,t)=e^{-i \epsilon(k)t} \Phi(k,t)
\ee
and using that $\Phi(k,t)=\Phi(k,t+T)$, where $T$ is the period ($\omega=2\pi/T$),
one gets that
\be
\left( H(k,t)-i\frac{\partial}{\partial t} \right) \Phi(k,t)=\epsilon(k) \Phi(k,t)
\label{gets}
\ee
The time-independent quasi-energies $\epsilon(k)$ are the eigenvalues of
the operator $H(k,t)-i\frac{\partial}{\partial t}$ and the function $\Phi(k,t)$ the
eigenfunction.
Note that due to the external time dependent perturbation, energy is not conserved
and therefore the original energy bands loose their meaning.
Since this function is periodic, we can expand it as
\be
\Phi(k,t)=\sum_m \phi_m(k) e^{i m \omega t}
\ee
Inserting this expansion in equation \ref{gets} we obtain the time-independent eigensystem
\be
\sum_{m^{\prime}} H_{m m^{\prime}}(k) \phi_{m^{\prime}} (k) = \epsilon(k) \phi_m (k)
\ee
with the quasi-energies the eigenvalues.
The Hamiltonian matrix is given by
\be
H_{m m^{\prime}}(k) = \delta_{m m^{\prime}} m \omega +
\frac{1}{T} \int_0^T dt e^{-i m \omega t} H(k,t) e^{i m^{\prime} \omega t}
\ee
Choosing a perturbation of the type $f(\omega t)=\cos (\omega t)$ the second term
of the Hamiltonian matrix reduces to $1/2 \left( \delta_{m^{\prime}+1,m} + \delta_{m^{\prime}-1,m} \right)$.

The time evolution of the state is then obtained solving for the quasi-energies, $\epsilon(k)$,
and the functions $\phi_m(k)$ diagonalizing the infinite matrix
\be
\label{hmm}
\left(\begin{array}{ccccc}
\cdots & \cdots & \cdots & \cdots & \cdots  \\
\cdots & (m-1)\omega + H(k) & \frac{1}{2} H_d(k)& 0 &\cdots \\
\cdots &  \frac{1}{2} H_d(k) & m\omega +H(k) & \frac{1}{2} H_d(k) &  \cdots \\
\cdots & 0 &   \frac{1}{2} H_d(k) & (m+1)\omega +H(k) & \cdots \\
\cdots & \cdots & \cdots & \cdots & \cdots
\end{array}\right)
\ee
The matrix can be reduced if the frequency is high enough and then only a few values of
$m$ are needed. 
In the case of a $2D$ triplet superconductor,
hopping, chemical potential, spin-orbit coupling
or magnetization, will be considered to vary with time. The first three parameters
preserve time reversal symmetry while the magnetization naturally breaks time
reversal symmetry if the unperturbed Hamiltonian is in a regime with vanishing
magnetization. Emphasis will be placed on the effects of varying the chemical potential
or the magnetization which are easilly tuned externally. In this last case it has
been determined before \cite{manisha} that even though the low energy states have a very low energy,
they may not be strictly Majorana fermions since the eigenvalues of the Floquet
operator (time evolution operator over one time period) are not strictly $\pm 1$.

Due to the periodicity of the eigenfunctions, $\Phi(k,t+T)=\Phi(k,t)$, the action of the evolution
operator, ${\cal U}(t)$, on a state over a period, $T$, leads to the same state minus a phase 
\be
|\psi(T)\rangle = {\cal U}(T) |\psi(0) \rangle = e^{-i \epsilon T} |\psi(0) \rangle
\ee
Therefore, the quasi-energies are defined minus a shift of a multiple of $w=2\pi/T$, and
we can restrict the quasi-energies to the first Floquet zone, defined by the interval
$-w/2\leq \epsilon \leq w/2$. States with quasi-energies $\epsilon=w/2$ and $\epsilon=-w/2$
are therefore equivalent and there is a reflection of any bands as one exits the Floquet
zone from above (or below) and as one enters from below (or above). Considering the particle-hole
symmetry of a superconductor, $\gamma_{-\epsilon}=\gamma_{\epsilon}^{\dagger}$ and the equivalence
between the energies $\epsilon=-w/2,w/2$ one expects a new type of finite quasi-energy Majorana mode in addition
to any zero energy states, the usual Majorana modes.

\begin{figure}[t]
\begin{center}
\includegraphics[width=0.48\textwidth]{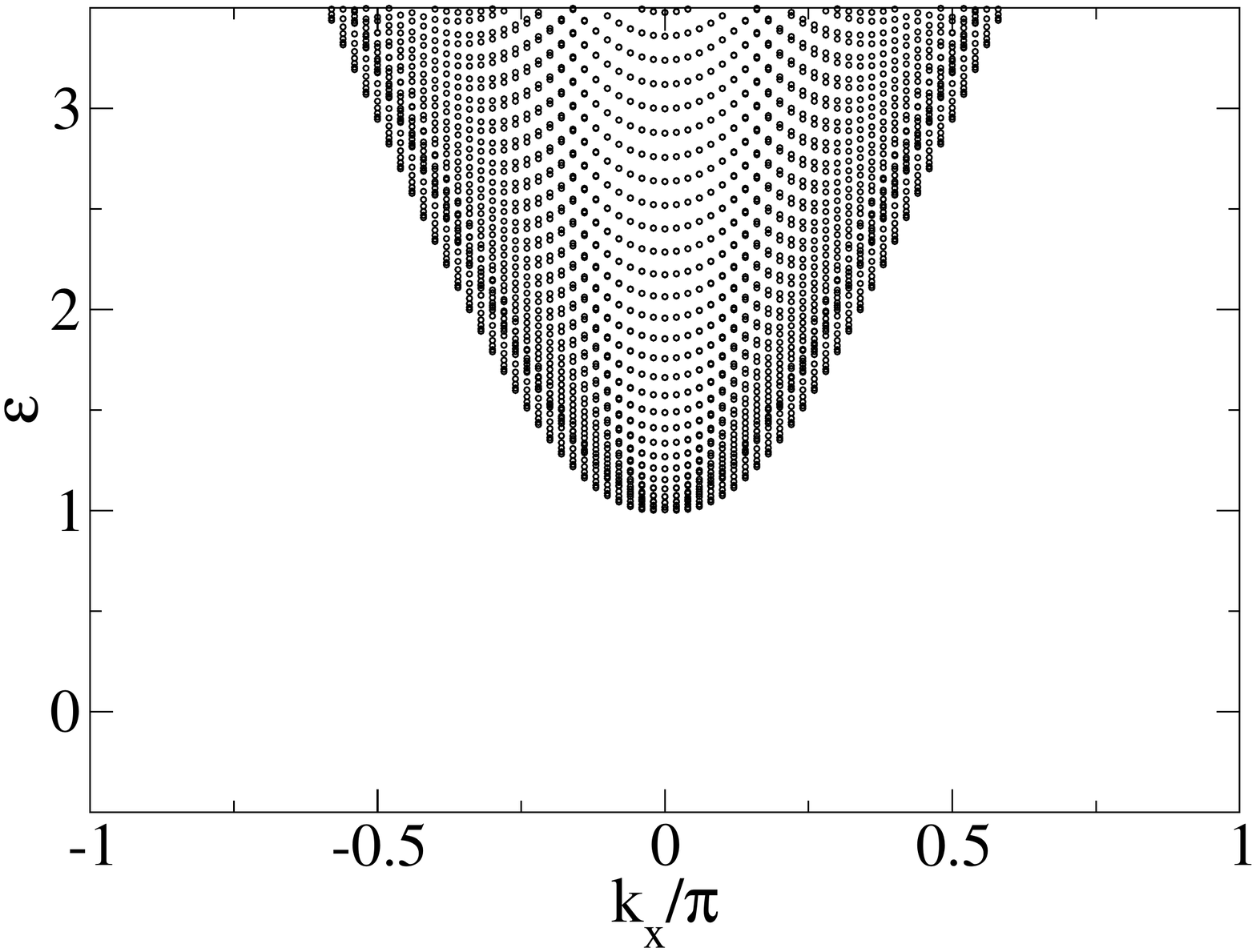}
\includegraphics[width=0.48\textwidth]{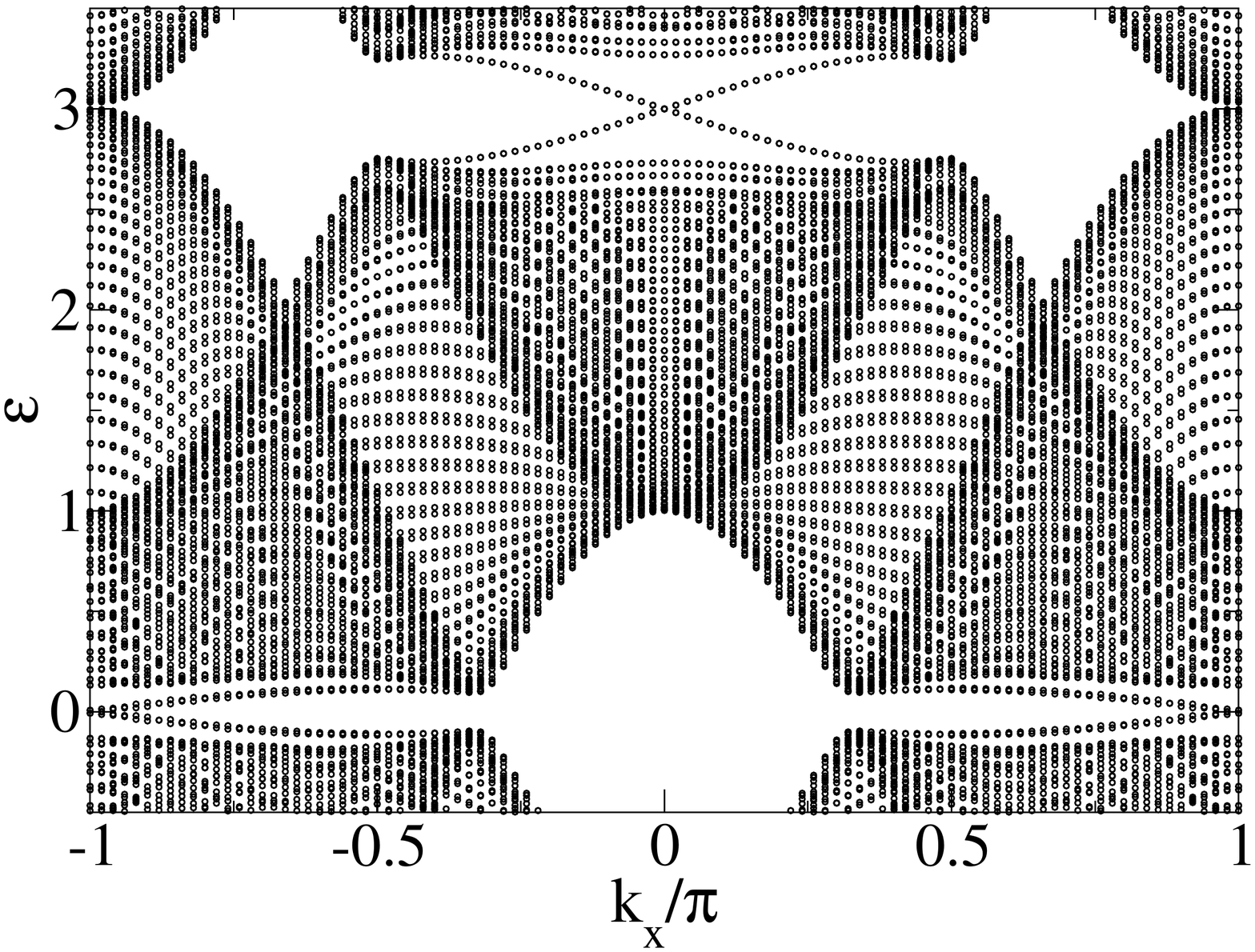}
\end{center}
\caption{\label{fig13}
Energy spectrum of (a) the unperturbed Hamiltonian (or Floquet spectrum for $l=0$) 
and (b) Floquet spectrum for $l=3$ for a $d_x,d_y$ triplet superconductor
in a topologically trivial phase
$\alpha=0.1,M_z=0,\mu=-5,\Delta_s=0.1,d_z=0,d=0.6$
 where the magnetization is changed
with time with frequency $w=6$. The periodic driving is $M_{zd} \cos wt$ with
$M_{zd}=4$. 
Reproduced from Ref. \cite{ref2}.
}
\end{figure}

\subsection{Quasi-energy bands of $2D$ triplet superconductor}

The solutions for the
quasi-energies of the perturbed Hamiltonians lead to bands that have a similar
structure to the energy bands of the unperturbed Hamiltonian obtained
taking a real space description with OBC along $y$ and a momentum description
along $x$. 

At large frequencies, $w>4\tilde{t}$, the size of the truncated matrix is relatively
small and the quasi-energies and physical properties (calculated over the
first Floquet zone) converge fast for small values of $m$.
Considering $m=0$ one reproduces the Hamiltonian of the unperturbed superconductor.
The first approximation for the driven system is obtained considering $m=1,0,-1$, then $m=2,1,0,-1,-2$ and so on.
One may therefore use a short notation for the number of terms considered in the diagonalization
of the Hamiltonian matrix by using $l=0,1,2,3,\cdots$. The unperturbed case is denoted by $l=0$
and the perturbed cases by $l=1,2,\cdots$ (considering that we are using $2l+1$ states). 
If the frequency $w$ is small, one needs to consider
large values of $l$ and the problem of finding the edge states in a ribbon geometry quickly
becomes heavy computationally. Increasing the value of the frequency it is easy to find that
it is enough to consider $l=2$, since taking $l=3$ leads to very similar results, with a good
accuracy. 

In Fig. \ref{fig13} we
consider periodic drivings in the magnetization for moderate
couplings of $M_{zd}=4$ and compare to the unperturbed case.
We consider frequency $w=6$.
In this case the unperturbed system is in a trivial phase evidenced by the
absence of gapless edge states inside the bulk gap.
Adding the perturbation edge states appear at low energies and also
appear at the border of the Floquet zone around $w/2$ (and $-w/2$). 
These states are also localized at the edges of the system.
In general, edge states appear at the border of the Floquet zone,
but as we can see from the figure there is no clearly defined gap throughout the
Brillouin zone. Edge states at low energies do not always appear or are mixed
with bulk edges. If the driving frequency is smaller or the perturbation has
a small amplitude the convergence is slow and in general the quasi-energy spectrum
is complex with a strong mixture of the edge and bulk states \cite{ref2}.

\subsection{Currents}

The edge states lead to the appearance of currents.
The charge current operator along direction $x$ at a given position $\hat{j}_y$ along $y$ is given by \cite{ref2}
\be
\hat{j}_c(j_y) = \frac{2e}{\hbar} \sum_{k_x} 
{\boldsymbol \psi}_{k_x,j_y}^{\dagger} 
\left(\begin{array}{cc}
-\tilde{t} \sin(k_x) & -\frac{i}{2} \alpha \cos (k_x) \\
\frac{i}{2} \alpha \cos(k_x) & -\tilde{t} \sin (k_x) 
\end{array} \right)
{\boldsymbol \psi}_{k_x,j_y} 
\ee
where ${\boldsymbol \psi}_{k_x,j_y}^{\dagger} =
\left( \psi_{k_x,j_y,\ua}^{\dagger}, \psi_{k_x,j_y,\da}^\dagger \right)$.
The current has contributions from the hopping and the spin-orbit terms.
The operators are written in real space along $y$ and in momentum space along $x$.
One may also define a longitudinal spin current, $\hat{j}_s(j_y)$, taking the difference between
the two diagonal components of the charge current. The other terms correspond to
spin-flip terms and do not contribute to the $z$ component of the spin current.

The average value of the charge current in the groundstate is given by summing over the
single particle occupied states (negative energies) in the usual way
\bea
j_c(j_y) = \langle \hat{j}_c(j_y) \rangle &=& \sum_{k_x,n} 
 \left\{ 
\tilde{t} \sin k_x \left[ 
  \tilde{v}_n(-k_x,j_y,\uparrow) \tilde{v}_n^*(-k_x.j_y,\uparrow) \right. \right.  
\nonumber \\
&+& \left. \tilde{v}_n(-k_x,j_y,\downarrow) \tilde{v}_n^*(-k_x.j_y,\downarrow) \right] \nonumber \\
&-& \frac{i \alpha}{2} \cos k_x \left[ 
 \tilde{v}_n(-k_x,j_y,\uparrow) \tilde{v}_n^*(-k_x.j_y,\downarrow) \right. \nonumber \\
&-& \left. \left. \tilde{v}_n(-k_x,j_y,\downarrow) \tilde{v}_n^*(-k_x.j_y,\uparrow)  \right]
\right\}
\eea
Here the functions are of the type
\be
\tilde{u}_n(k_x,j_y,\sigma) = \sum_m e^{imwt} u_{n,m}(k_x,j_y,\sigma)
\ee
where as usual $\sigma=\uparrow,\downarrow$.

\begin{figure}[t]
\begin{center}
\includegraphics[width=0.7\textwidth]{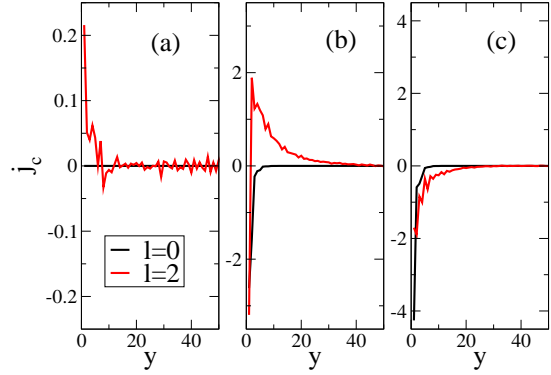}
\end{center}
\caption{\label{fig14}
(Color online)
Charge current profiles for the unperturbed ($l=0$) and
perturbed ($l=2$) $d_x,d_y$ superconductor
for  (a) $\mu=-5,M_z=0$ (with $C=0$), 
(b) $\mu=-5,M_z=2$ (with $C=1$) and
(c) $\mu=-1,M_z=2$ (with $C=-2$) and
for $\mu_d=1$.
Only one half of the system is shown since the current profile is anti-symmetric
around the middle point. The system size is $N_y=100$.
}
\end{figure}

The charge currents at the edges of the unperturbed Hamiltonian are well
understood. If there is TRS the currents vanish and if TRS is broken the
edge charge currents are finite (finite Chern number). We consider as example
a $d_x,d_y$ triplet superconductor. 
If the magnetization vanishes, the system has TRS and vanishing
charge edge currents in the topologically trivial phases. The charge current
also vanishes in the $Z_2$ topological
phase but the spin edge currents are non-vanishing.

We consider a set of parameters
$d=0.6,\Delta_s=0.1,d_z=0,\alpha=0.6$ and different values for the chemical
potential and the magnetization. 
In Fig. \ref{fig14} we show results for the profile of the charge current as a function of $y$
for various cases. We compare the unperturbed case with the perturbed one
by considering that at $w=6$ it is enough to truncate the
Hamiltonian matrix at $l=2$. To calculate the currents we sum over the states
in the first Floquet zone. Also the results are for time $t=0$ or any multiple
of the time period $T$.
As seen in Fig. \ref{fig14}a the periodic driving gives rise to a finite
charge current that is absent in the unperturbed trivial phase. In the other
two panels the edge states of the unperturbed system carry a finite current which is altered by
the periodic driving due to the appearance of extra edge states and also a
reshaping of the continuum states; notably there is a change of sign in Fig. \ref{fig14}b.

As shown in Ref. \cite{ref2} the edge states also generate spin currents.
Associated with the spin currents in the topological phases, it has been
shown that the eigenstates have non trivial spin polarizations that 
depend strongly on the momentum \cite{timm}.  
In Ref. \cite{ref2} the spin polarization of the induced edge states is also
considered.

\section{Conclusions}

In this work the robustness of edge modes of topological systems to time-dependent perturbations
was considered.
The fermionic and Majorana edge mode dynamics of various topological systems were
compared, after a global quench of the Hamiltonian parameters takes place. Also,
the effect of a periodic perturbation was considered.
In general the edge modes are not stable, however in finite systems there
is the possibility of revival after a finite time. It was shown that 
the distinction due to the Majorana nature of the excitations plays a small role in
comparison to the details of the energy spectrum and overlaps between states.

Slow transformations were not considered here but allow the occurence of the
Kibble-Zurek mechanism of defect production as one crosses
a quantum critical point.
In the context
of the Creutz ladder, it was shown before that the presence of edge states modifies the
process of defect production expected from the Kibble-Zurek mechanism, leading
in this problem to a scaling with the change rate with a non-universal critical
exponent \cite{bermudez1}. A similar result was obtained for the one-dimensional
superconducting Kitaev model, where it was shown that, although bulk states follow
the Kibble-Zurek scaling, the produced defects for an edge state quench are quite
anomalous and independent of the quench rate \cite{bermudez2}.
Similar results have been found for a $2D$ triplet superconductor \cite{ref1}.
As in the case of sudden quenches, there seems to be no particular signature of
the Majorana fermions in comparison to other edge modes. Note that Majoranas 
are absent in the Creutz ladder.

Fermionic edge states
in a topological insulator are now established \cite{bhz}. 
Even though Majorana edge states have been extensively studied in the
literature their experimental detection has proved challenging. 
While there is
promising evidence of Majorana edge states \cite{mourik,science}
in magnetic chains superimposed on a conventional superconductor,
there may be other sources of the edge states observed in the system
considered (see for instance ref. \cite{dumitrescu} for
a discussion and references therein).
 
One of the methods proposed to detect the presence of Majorana edge states is
the measurement of the differential conductance at the interface between a lead
and a topological superconductor.
If the lead is metallic one expects a zero-bias peak in the differential
conductance, if zero-energy modes are present in the superconducting side.
In the presence of Majorana modes one expects a vanishing conductance if the number of
Majorana modes is even and a quantized value of $2e^2/h$, if the number of modes is odd \cite{law,wimmer}.
In the case of the dimerized SSH model here considered,
it has been shown \cite{tanaka} that the fermionic edge modes do not contribute to the
conductance and, therefore, provides a method to distinguish the various phases and
the type of edge modes.

\section*{Acknowledgements}
The author acknowledges discussions with Pedro Ribeiro, Antonio Garcia-Garcia,
Xiaosen Yang, Maxim Dzero and Henrik Johannesson. 
Partial support in the form of a BEV by the CNPq at CBPF (Rio de Janeiro) 
and partial support and hospitality by the Department of Physics of 
Gothenburg University are acknowledged.
Partial support from the 
Portuguese FCT under Grants No. PEST- OE/FIS/UI0091/2011, No. PTDC/FIS/111348/2009
and UID/CTM/04540/2013 is also acknowledged.

\end{document}